\documentclass[useAMS,usenatbib]{mn2e}
\usepackage{graphicx}
\usepackage{rotating,times,pictex,graphicx,latexsym}
\usepackage{color}
\usepackage{longtable,amsmath}
\usepackage{lscape}
\usepackage{array}
\usepackage{lipsum}
\newcolumntype{C}[1]{>{\centering\arraybackslash}p{1.1cm}}

\newcommand{\htwo}{H$_2$}

\newcommand{\hii}{H{\sc ii}}

\title[YSO jets in the Galactic Plane]{YSO jets in the Galactic Plane from
UWISH2:\\ III - Jets and Outflows in Cassiopeia and Auriga} 

\author[Froebrich et al.]{D.~Froebrich$^{1}$\thanks{E-mail: df@star.kent.ac.uk},
S.V.~Makin$^{1}$\\ $^1$ Centre for Astrophysics and Planetary Science, University
of Kent, Canterbury, CT2 7NH, UK } 

\begin{document}

\date{Received sooner; accepted later}
\pagerange{\pageref{firstpage}--\pageref{lastpage}} \pubyear{2011}
\maketitle

\label{firstpage}

\begin{abstract}

We present the analysis of 35.5 square degrees of images in the 1\,--\,0\,S(1)
line of \htwo\ from the UK Widefield Infrared Survey for \htwo\ (UWISH2) towards
Cassiopeia and Auriga. We have identified 98 Molecular Hydrogen emission-line
Objects (MHOs) driven by Young Stellar Objects, 60\,\% of which are bipolar
outflows and all are new discoveries. We estimate that the UWISH2 extended
emission object catalogue contains fewer than 2\,\% false positives and is
complete at the 95\,\% level for jets and outflows brighter than the UWISH2
detection limit. We identified reliable driving source candidates for three
quarters of the detected outflows, 40\,\% of which are associated with groups
and clusters of stars. The driving source candidates are 20\,\% protostars, the
remainder are CTTSs. We also identified 15 new star cluster candidates near MHOs
in the survey area.

We find that the typical outflow identified in the sample has the following
characteristics: the position angles are randomly orientated; bipolar outflows
are straight within a few degrees; the two lobes are slightly asymmetrical in
length and brightness; the length and brightness of the lobes are not
correlated; typical time gaps between major ejections of material are
1\,--\,3\,kyr, hence FU-Ori or EX-Ori eruptions are most likely not the cause of
these, but we suggest MNors as a possible source. Furthermore, we find that
outflow lobe length distributions are statistically different from the widely
used total length distributions. There are a larger than expected number of
bright outflows indicating that the flux distribution does not follow a power
law.

\end{abstract}


\section{Introduction}

The formation of stars via disk accretion of material is inevitably related to
mass ejection into jets and outflows along the rotational axis of these objects.
The outflows from protostars and Young Stellar Objects (YSOs) were first
correctly recognised as such by \citet{1980ApJ...239L..17S}. Since then numerous
studies of these jets and outflows have been conducted to investigate the
details of the excitation mechanism, the mass ejection rates, the jet launching,
acceleration and collimation, as well as the outflow energetics and timescales
(see reviews of e.g. \citet{1996ARA&A..34..111B}, \citet{2000prpl.conf..867R},
\citet{2007prpl.conf..215B}, \citet{2014prpl.conf..451F}). However, there are
still a number of unsolved questions about the outflow phenomenon. Which factors
statistically determine the properties (length, luminosity, formation of the
main \htwo\ emission features -- knots) of the outflows? Are the properties of
the central source (final mass, age) responsible for these or has the
environment (density structure in low mass vs. high mass star forming regions) a
significant influence? Is the energy and momentum feedback  from the outflows
significant enough to explain the local turbulent energy near the forming stars
and are the jets and outflows able to terminate the star formation process
locally? In order to answer these and similar questions, we need to
observationally establish the number of jets and outflows from young stars in
the Galactic plane and to determine their average properties. 

The first truly large scale work to identify all jets and outflows in a star
forming region via an unbiased survey was done in Orion\,A by
\citet{Stanke2002}. They utilised the molecular hydrogen ro-vibrational
1\,--\,0\,S(1) transition at 2.122\,$\mu$m. This line is a proven excellent
tracer of hot (T$\sim$2000\,K) and dense (n$\ge$10$^3$\,cm$^{-3}$) gas excited
by the fast shocks ($10 - 100$\,km\,s$^{-1}$) caused by the interactions of jets
and outflows with the surrounding interstellar medium. As the line is in the
K-band, it is less influenced by extinction compared to other tracers of these
shocks such as optical H$\alpha$ or [SII] lines, which are the historically used
tracers for Herbig-Haro objects. It is usually also stronger than the near
infrared (NIR) [FeII] lines, except in strong shocks or purely atomic
environments. Observations in other molecular outflow tracers, such as CO or
SiO, on large (molecular cloud) scales typically lack the combination of spatial
resolution and depth to identify the fainter outflows, which are detectable in
the NIR, especially in complex regions along the Galactic plane. Furthermore,
the 1\,--\,0\,S(1)  line flux is proportional to the total outflow luminosity
for a range of excitation conditions \citep{CarattioGaratti2006}. Note that
despite the shock velocity and gas density limitations for the excitation of the
1\,--\,0\,S(1) transition, at least some parts of the vast majority of jets and
outflows from YSOs are detectable in this line.

After the pioneering work by \citet{Stanke2002}, further searches for jets and
outflows in star forming regions have been conducted e.g. by
\citet{Walawender2005}, \citet{Hatchell2007}, \citet{Davis2009} and
\citet{Khanzadyan2012}. In order to establish a truly unbiased sample of jets
and outflows from young stars in the Galaxy, not restricted to nearby, mostly
low-mass star forming regions, the UKIRT Widefield Infrared Survey for \htwo\
(UWISH2) was conducted by \citet{Froebrich2011}. In this series of papers we are
analysing in detail the identified jets and outflows from young stars in this
survey as Molecular Hydrogen emission-line Objects (MHOs) defined in 
\citet{2010A&A...511A..24D}. Previous works based on UWISH2 data by
\citet{Ioannidis2012,2012MNRAS.425.1380I} have concentrated on the Serpens and
Aquila region along the plane, while
\citet{2012ApJS..200....2L,2013ApJS..208...23L} have investigated \htwo\ ouflows
from Spitzer-detected extended green objects. 

This paper is structured as follows: In Sect.\,\ref{dataandanalysis} we briefly
discuss the data and our analysis procedures such as the identification of the
MHOs and and their most likely driving sources. The results, outflow and driving
source properties, are then discussed in detail in Sect.\,\ref{results}.

\section{Data analysis}\label{dataandanalysis}

\subsection{UWISH2 data}

Our analysis uses data from the UKIRT Widefield Infrared Survey for \htwo\ by
\citet{Froebrich2011} and its extension towards the Cygnus\,X and
Cassiopeia/Auriga regions discussed in \citet{2015MNRAS.454.2586F}. All images
are taken with the Wide Field Camera \citep{Casali2007} at the UK Infra-Red
Telescope. A total exposure time of 720\,s per pixel was obtained through the
1\,--\,0\,S(1) filter at 2.122\,$\mu$m. Utilising micro-stepping during the
observations, the final images have a pixel size of
0.2\arcsec$\times$0.2\arcsec\ and the typical seeing in the data is 0.8\arcsec.
Averaged over this typical seeing, the typical 5$\sigma$ surface brightness
detection limit in the data is
4.1\,$\times$\,10$^{-19}$\,W\,m$^{-2}$\,arcsec$^{-2}$. 

\begin{figure}
\centering
\includegraphics[angle=0,width=\columnwidth]{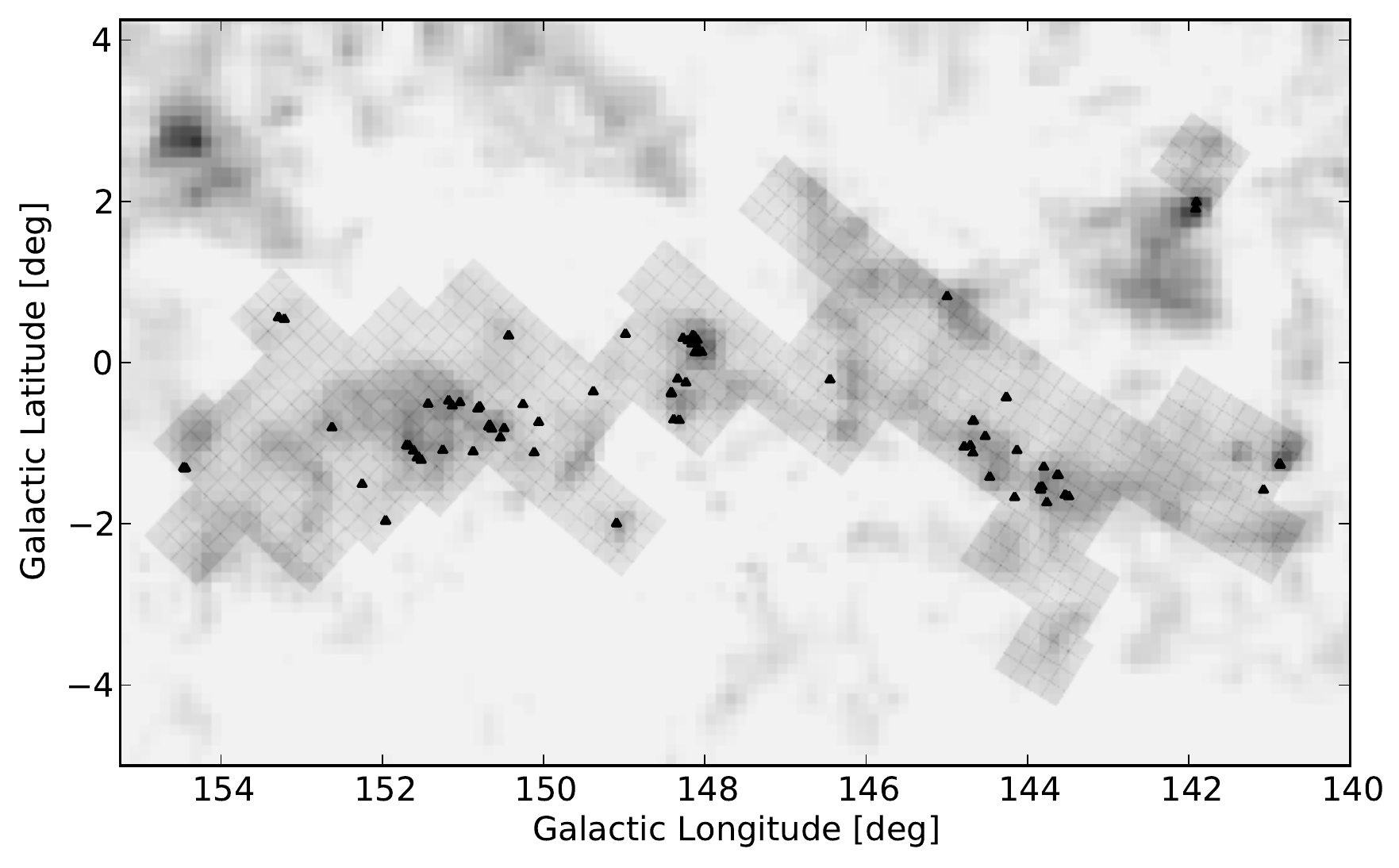} 

\caption{Coverage of the survey along the Galactic plane in Cassiopeia and
Auriga with UWISH2 tiles in Galactic Coordinates. Observed fields are indicated
by the slightly darker grid regions. Overplotted as black triangles are the
positions of all detected MHOs in this area. The background image are the CO
data from \citet{2001ApJ...547..792D}; darker colours indicate higher levels of
CO emission. \label{coverage_figure}}

\end{figure}

In this paper we only analyse the data taken towards the region of
Cassiopeia and Auriga. This covers Galactic Longitudes from approximately $l =
140\degr$ to $l = 155\degr$ and Galactic Latitudes from $b = -4\degr$ to $b =
3\degr$. The detailed coverage of the field is shown in
Fig.\,\ref{coverage_figure}. The observed fields cover a total of 35.46 square
degrees. For the entire area we also have broad-band JHK data from the UKIDSS
\citep{Lawrence2007} Galactic Plane Survey \citep{Lucas2008}. The K-band images
have been utilised to continuum subtract the \htwo\ images following the
procedures described in \citet{2014MNRAS.443.2650L}.

\subsection{MHO and driving source identification}

The identification of jets and outflows from young stars, or MHOs -- as defined
in \citet{2010A&A...511A..24D}, and their respective driving sources has been
done in the following way:

\begin{itemize}

\item We selected all \htwo\ emission line features that were catalogued by
\citet{2015MNRAS.454.2586F} in the Cassiopeia and Auriga region of UWISH2.
There were 51 groups of jets features with a total of 187 individual \htwo\
emission line objects in this area. These groups were originally defined in
\citet{2015MNRAS.454.2586F} only based on their spatial distribution. Here we
set out to identify the individual outflows in each of these groups. 

\item Each of the groups was investigated in detail by inspecting the \htwo-K
difference images and JK\htwo\ colour composite images. We tried to assign the
individual \htwo\ features in each group to as few as possible jets/outflows.
The decision of which \htwo\ emission line object belongs to which potential
outflow has been based on the appearance/shape of the emission feature, and the
alignment of features with each other and/or the potential driving source. Each
identified outflow has subsequently been assigned a unique MHO number. \htwo\
features (or small groups thereof) without an obvious driving source candidate
have also been given a unique MHO number if their appearance suggests they
represent emission from shocked gas. \htwo\ features that are most likely caused
by radiative excitation/fluorescence have been disregarded from any further
analysis.

\item We also searched the \htwo-K and JK\htwo\ images near the pre-selected
groups of jet features for objects that have been missed in the original
catalogue. These objects were missing as they did not meet the selection
criteria in \citet{2015MNRAS.454.2586F}. Most of the identified outflows have
one or a few \htwo\ emission knots that have a surface brightness that prevented
their inclusion in the original list.

\item As MHO coordinates we use the coordinates of the candidate driving source
if one can be identified. For MHOs consisting of single or multiple knots
without source candidate, the mean coordinates of the knots are used as MHO
position.

\item Driving source candidates were identified based mainly on their alignment
with the potential outflow direction. Furthermore, we considered the NIR or
mid-IR colours/detections of potential source candidates as well as the K-band
variability and the presence of nebulosities near potential sources. Based on
these criteria and the resulting number of potential driving sources nearby, we
assigned a likelihood to each identified source that indicates how certain we
are that the correct source has been identified for the \htwo\ features. In
some cases the position of the suspected driving source is clear, but there is
no detection in any catalogue or the literature. In these cases we have
manually determined the suspected source position and indicated this scenario
with $noname$ as the source identification for the MHO.

\end{itemize}

\subsection{Determination of MHO and driving source properties}

After the identification of the individual MHOs as outflows or groups of
individual \htwo\ features, we grouped the MHOs into a number of categories and
determined the properties for each MHO as well as those of the driving sources.
The details of this process are as follows:

\begin{itemize}

\item MHOs are categorised as either bipolar outflows, single-sided outflows or
just an individual/small group of \htwo\ emission knots.

\item For all MHOs, sub-features are identified and labelled by letters. We
start at one end of the outflow with 'A' and knots are labelled with sequential
letters along the outflow axis. For groups of \htwo\ features without a source,
there is no systematic pre-described order in which the labelling of the
sub-features is done. 

\item For each identified outflow lobe, the length of the lobe from the
potential source to the most distant end of the \htwo\ knot is measured, as well
as the position angle of the lobe with respect to the North direction. For all
bipolar outflows, the difference of the two position angles from a perfect
parallel orientation of the two lobes is calculated.

\item For each outflow lobe with more than one \htwo\ emission knot, all the 
gaps/distances between succeeding knots are measured. 

\item Photometry is done separately for each \htwo\ knot that has been assigned
to any of the MHOs. All knots are outlined with polygons enclosing the area
considered to be part of the \htwo\ emission feature and the total integrated
flux inside the polygon is measured. Care has been taken to exclude any
continuum point sources superimposed on the \htwo\ emission from the flux
measurements, and also low surface brightness areas are included, which were not
picked up in the automatic \htwo\ feature detection used in
\citet{2015MNRAS.454.2586F}. For each MHO a nearby, \htwo\ emission-free area of
the image has been selected to estimate the 'local' background flux. The
calibration of the fluxes has been done in the same way as described in
\citet{2015MNRAS.454.2586F}.

\item For each identified MHO driving source we estimate the likelihood that we
have selected the correct source, based on the number of nearby potential source
candidates and their properties. We also categorise each MHO according to
whether it is likely to come from a driving source situated in a cluster or
group of stars, or if the potential source is isolated.

\item For each driving source the coordinates and designations (e.g. 2MASS, 
WISE, or IRAS IDs) are collected. If a source is detected in multiple
catalogues, we use the coordinates of the object in the catalogue with the
shortest wavelengths, as these usually have the highest spatial resolution.

\item For each driving source the near- and mid-IR magnitudes from 2MASS, GPS and
WISE are collected. We determine the K-band variability for all objects that
have more than one K-band magnitude available. 

\item We utilise the WISE magnitudes to determine the spectral index $\alpha$ of
the SED following \citet{2013Ap&SS.344..175M}, where positive values denote
colder objects and negative values warmer objects. We use the $\alpha$ value to
sort the driving source candidates with WISE detections into an age sequence.
Hereafter we split the sample into younger and older objects at $\alpha = -0.5$,
which ensures a 50/50 split of the objects into the two groups. Note that the
younger sample will hence contain protostars and CTTSs, i.e. the two samples do
not represent a split into the two evolutionary stages.

\end{itemize}

\section{Results and Discussion}\label{results}

\subsection{The Cassiopeia and Auriga MHO catalogue}

From the investigated 51 groups of \htwo\ features, 46 (90\,\%) are found to
contain emission knots which most likely are caused by shock excitation from
jets and outflows from young stars. Two of the five groups without MHOs are
most likely caused by fluorescently excited molecular cloud edges. Two further
objects are most likely variable stars (note that the \htwo\ and K-band images
are taken up to several years apart), and one object seems to be an image
artefact from a bright star. In total there are 15 (8\,\%) individual \htwo\
features, originally classified as 'jet' in \citet{2015MNRAS.454.2586F} that
after detailed investigation have been judged not to be shock excited emission.
The majority (12 or 6.4\,\%) are most likely associated with fluorescently
excited molecular cloud edges. Only 3 (1.6\,\%) of the \htwo\ features are
artefacts or variable stars. Thus, even given the small sample size, one can
conclude that the full UWISH2 catalogue from \citet{2015MNRAS.454.2586F}
contains fewer than 2\,\% false positive detections of \htwo\ features, and the
fraction of misclassified 'jet' features in the catalogue is well below 10\,\%.

In the 46 groups with jet/outflow-like features we have identified 98 individual
MHOs. Thus, the average number of outflows per jet group is about two. Hence,
scaled up to the full UWISH2 catalogue there will be about 1500 outflows from
young stars. We have overplotted the positions of all MHOs on the CO data from
\citet{2001ApJ...547..792D} in Fig.\,\ref{coverage_figure}. As expected, the
objects are mostly concentrated towards the high column density CO features.
They are also clustered in groups of a few, similar to what has been found in
previous studies along the Galactic plane in Serpens and Aquila by
\citet{Ioannidis2012}.

There are 57 (58\,\%) bipolar and 17 (17\,\%) single-sided outflows amongst the
98 MHOs. For the remaining 24 (24\,\%) MHOs, which are groups of or individual
\htwo\ knots, no convincing source candidate could be identified. Driving source
candidates have been identified for all the bipolar and single-sided objects.
Note that all MHOs in this regions are new discoveries. Potentially the only
known outflow might be MHO\,2982 with the source candidate AFGL\,490-iki
(IRAS\,03234+5843) which has been discovered as a roughly North-South orientated
CO-outflow in \citet{1998AJ....116..840L}. The \htwo\ detection in our data
coincides with the redshifted lobe of the CO data.

Many of the MHOs have \htwo\ emission knots which are not included in the full
UWISH2 catalogue. However, only four of the MHOs (4\,\%) had none of their
\htwo\  knots in the catalogue. Thus, the list of \citet{2015MNRAS.454.2586F}
has a completeness of the order of 95\,\% or higher for \htwo\ emitting jets and
outflows from young stars brighter than the UWISH detection limit.

In Table\,\ref{datatable} we list all the properties of the identified MHOs.
This includes the following: i) the MHO number; ii) Right Ascension of the
object; iii) Declination of the object; iv) Length of the outflow lobe(s); v)
Position angle of the outflow lobe(s) from North over East; vi) Flux of the
outflow lobe(s) or knots; vii) Type of outflow (B - bipolar, S - single-sided, K
- single or group of knots without apparent source candidate); viii) Is the MHO
associated with cluster or group of stars? ix) Source candidate identification;
x) Likelihood P$_S$ that source candidate is the real driving source; 
xi)\,--\,xiii) Near-infrared JHK magnitudes of the source candidate;
xiv)\,--\,xvii) Mid-infrared WISE magnitudes of the source candidate;
xviii)\,--\,xxii) Is there a detection of the source candidate in the following
surveys/catalogues: G -- UKIDSS GPS, 2 -- 2MASS, W -- WISE, A -- AKARI, I --
IRAS? In the Appendix table we also show the \htwo-K difference and JK\htwo\
colour composite images for all MHOs together with some additional notes on each
object.

\subsection{The driving source properties}

We have identified a driving source candidate for 74 (76\,\%) of the MHOs, i.e.
for all the single-sided and bipolar outflows. Note that some MHOs have the same
source candidate (e.g. MHO\,1070, 1071, 1072). Even if we could not find a
convincing source candidate for the remaining MHO objects, it is possible to
judge whether the driving source is likely to be in a group or cluster of stars,
or isolated. In total 40 (41\,\%) of the potential driving sources are
associated with clusters or groups of stars, while 58 (59\,\%) seem to be
isolated. As the majority of the young protostars ($\alpha > 0$) and CTTSs
($\alpha <0$) are situated in clusters a larger fraction of potential driving
sources are expected to be associated with clusters. A similar result has been
found for outflows in Orion\,A by \citet{Davis2009}. This seems to indicate that
either clustered star formation inhibits the formation of outflows or that due
to the shortness and early onset (a larger fraction of protostars drive outflows
compared to CTTSs) of the outflow active state, the sources in clusters
identifiable in the NIR data are simply older than the isolated sources. Thus,
the observed percentages could simply reflect an evolutionary trend. This is
somewhat supported by the fact that 28\,\% of sources in clusters are in the
younger group, while this is the case for 41\,\% of the isolated sources.

In total we have identified 15 as-yet unknown cluster candidates and seven
apparent groups of YSOs near the MHOs in the survey area. Several of these new
clusters are compact and associated with IRAS sources. We list their positions,
apparent radii and the estimated number of cluster members visible in the NIR
images in Table\,\ref{clustertable}. Our distinction of 'clusters' and 'groups'
has been based on the number of potential members as well as the distribution
of them. Where the distribution appears circular the object is classified as
cluster even if the number of visible members is low. If the number of members
is low and they seem to be following a more 'filamentary' distribution the
object is classified as a group. Note that we have not performed an exhaustive,
complete search for clusters in the entire survey area discussed here. Thus,
the number of so far undiscovered clusters in the survey area is potentially
much higher.

\begin{figure}
\centering
\includegraphics[angle=0,width=\columnwidth]{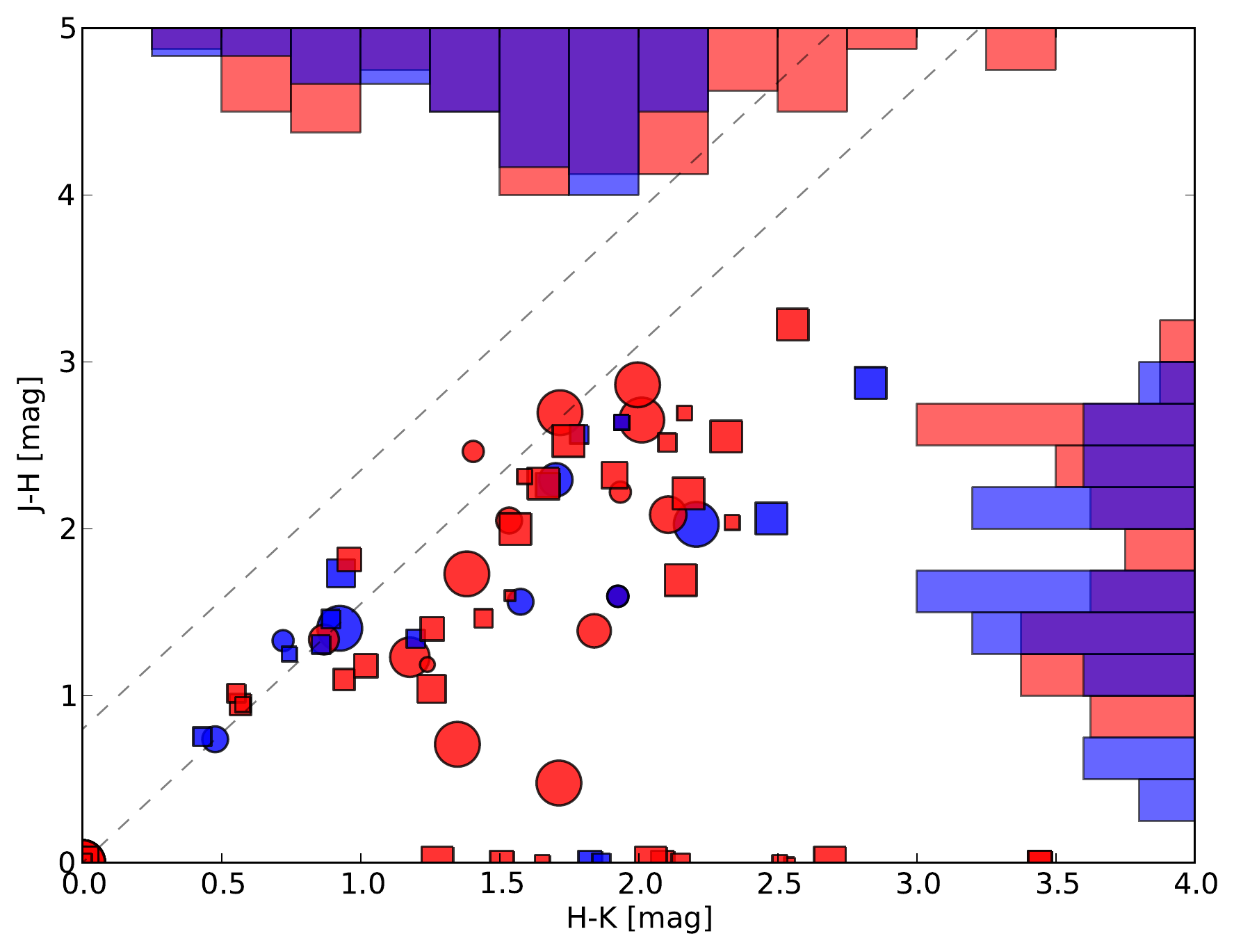} 

\caption{NIR colour-colour-diagram of the identified candidate outflow driving sources. The dashed
lines indicate the standard reddening band for stellar atmospheres. Square
symbols indicate GPS detections, while circular symbols indicate 2MASS data. Red
(lighter) symbols are for bipolar outflows, while the blue (darker) symbols are
for single-sided MHOs. The size of the symbols indicates the probability
assigned to each source candidate that the identification is correct. The
histograms show the NIR colour distribution of the source candidates (red/light
for GPS and blue/dark for 2MASS). \label{ccd_figure}}

\end{figure}

Of all driving source candidates 59 (80\,\%) have a WISE detection in all four
filters. The classification of the objects based on the WISE fluxes following
\citep{2013Ap&SS.344..175M} shows that 10 (17\,\%) have positive slopes
($\alpha$), while 49 (83\,\%) have a negative slope of the SED, indicating a
mix of CTTSs and protostars as the driving sources for the detected outflows.
We also investigated the NIR colours of the sources. For 23 (31\,\%) objects we
found a 2MASS counterpart, and 35 (47\,\%) are detected in GPS.
Figure\,\ref{ccd_figure} shows the NIR colour-colour diagram of all the source
candidates. Most objects are below or at the bottom of the reddening band,
indicating CTTSs and protostars as driving sources.

In total 23 (31\,\%) of the source candidates have more than one K-band
detection (i.e. GPS and 2MASS) and we investigate their K-band variability
between the two surveys. Of these objects, 16 (70\,\%) show a variability of
more than 0.1\,mag, and 4 (17\,\%) vary by more than 0.5\,mag over a typical
timescale of several years. Only 4 objects have two epochs of observations in
GPS, none of them varies by more than 0.1\,mag between these two epochs.

\begin{table*}

\caption{Summary table listing the properties of the newly discovered clusters
and groups of stars near the MHOs. We list: i) The cluster ID; ii) The cluster
Right Ascension (J2000) in degrees; iii) The cluster Declination (J2000) in
degrees; iv) The apparent cluster radius in arcminutes; v) C -- object most
likely a cluster; G -- object most likely just a small group of stars; vi)
Estimated number of near-infrared visible members; vii) ID numbers of MHOs with
driving sources in the cluster; viii) Notes on each object such as the
association with known objects. \label{clustertable} }

\renewcommand{\tabcolsep}{3pt}

\begin{center}
\begin{tabular}{c|c|c|c|c|c|c|p{9cm}}
\hline
ID & RA [deg] & DEC [deg] & R ['] & C/G & Stars & MHO & Notes\\
\hline
01 & 51.111 & 55.204 & 0.35 & C & 50 & 2986,2987,2988 & coincides with IRAS\,03205+5501 \\
02 & 61.124 & 51.404 & 0.65 & C & 40 &                & no known associations \\
03 & 60.831 & 51.487 & 0.77 & C & 30 & 1068           & no known associations \\
04 & 61.880 & 50.513 & 0.40 & C & 30 &                & coincides with IRAS\,04037+5022 \\
05 & 50.993 & 55.186 & 0.46 & C & 30 &                & no known associations \\
06 & 61.638 & 50.508 & 0.55 & C & 20 & 1087           & no known associations \\
07 & 62.061 & 50.519 & 1.24 & C & 20 & 1090,1091      & coincides with IRAS\,04045+5023, very extended \\
08 & 52.990 & 55.646 & 0.27 & C & 20 & 2996,2997      & coincides with IRAS\,03281+5528 \\
09 & 61.362 & 49.652 & 0.37 & C & 15 & 1092,1093,1094 & 1' north of \hii\ region, classified as Galaxy (LEDA\,2348913) \\
10 & 63.340 & 50.056 & 0.23 & C & 15 & 1096           & no known associations \\
11 & 61.486 & 51.451 & 0.19 & C & 10 & 1078           & dominated by bright nebulous star 2MASX\,J04055657+5127052 (Galaxy) \\
12 & 65.569 & 50.572 & 0.25 & C & 10 & 1097,1098      & no known associations \\
13 & 65.468 & 50.610 & 0.47 & C & 10 & 1099           & coincides with IRAS\,04181+5029 \\
14 & 53.289 & 55.174 & 0.41 & C & 10 & 3700,3701,3702 & 0.5' South of \hii\ region MSX6C\,G144.6678-00.7136 \\
15 & 56.295 & 54.528 & 0.25 & C & 10 & 3707           & 0.5' East of IRAS\,03412+5422 \\ \hline
16 & 61.090 & 51.396 & 0.40 & G & 15 & 1073,1074      & near \hii\ region, classified as Galaxy (2MASX\,J04041342+5122587) \\
17 & 61.955 & 51.257 & 1.13 & G & 15 & 1081           & coincides with IRAS\,04040+5107 \\
18 & 62.015 & 50.492 & 0.39 & G & 15 &                & no known associations \\
19 & 61.013 & 51.378 & 0.32 & G & 10 & 1070,1071,1072 & about 8' from large \hii\ region NGC\,1491 \\
20 & 61.731 & 50.500 & 0.20 & G & 10 & 1085           & no known associations \\
21 & 50.659 & 55.061 & 0.18 & G & 10 & 2984,2985      & coincides with IRAS\,03188+5452 \\
22 & 51.617 & 54.682 & 0.08 & G &  5 & 2995           & coincides with IRAS\,03226+5430 \\
\hline
\end{tabular}
\end{center}
\end{table*}

\subsection{The MHO properties}

40 MHOs have potential source candidates inside or associated with (nearby,
potentially triggered SF) young clusters, while 58 MHOs have potential sources
with no apparent connection to a clustered environment. Of the bipolar objects,
24 (42\,\%) are associated with clusters and 33 (58\,\%) are not. Hence the
visibility of one or two of the outflow lobes does not depend on the
environment. Note that "clustered environment" in this region typically refers
to clusters or groups with 10\,--\,50 NIR visible stars and not massive clusters
(with a few exceptions).

\subsubsection{Jet/Outflow orientation}

\begin{figure}
\centering
\includegraphics[angle=0,width=\columnwidth]{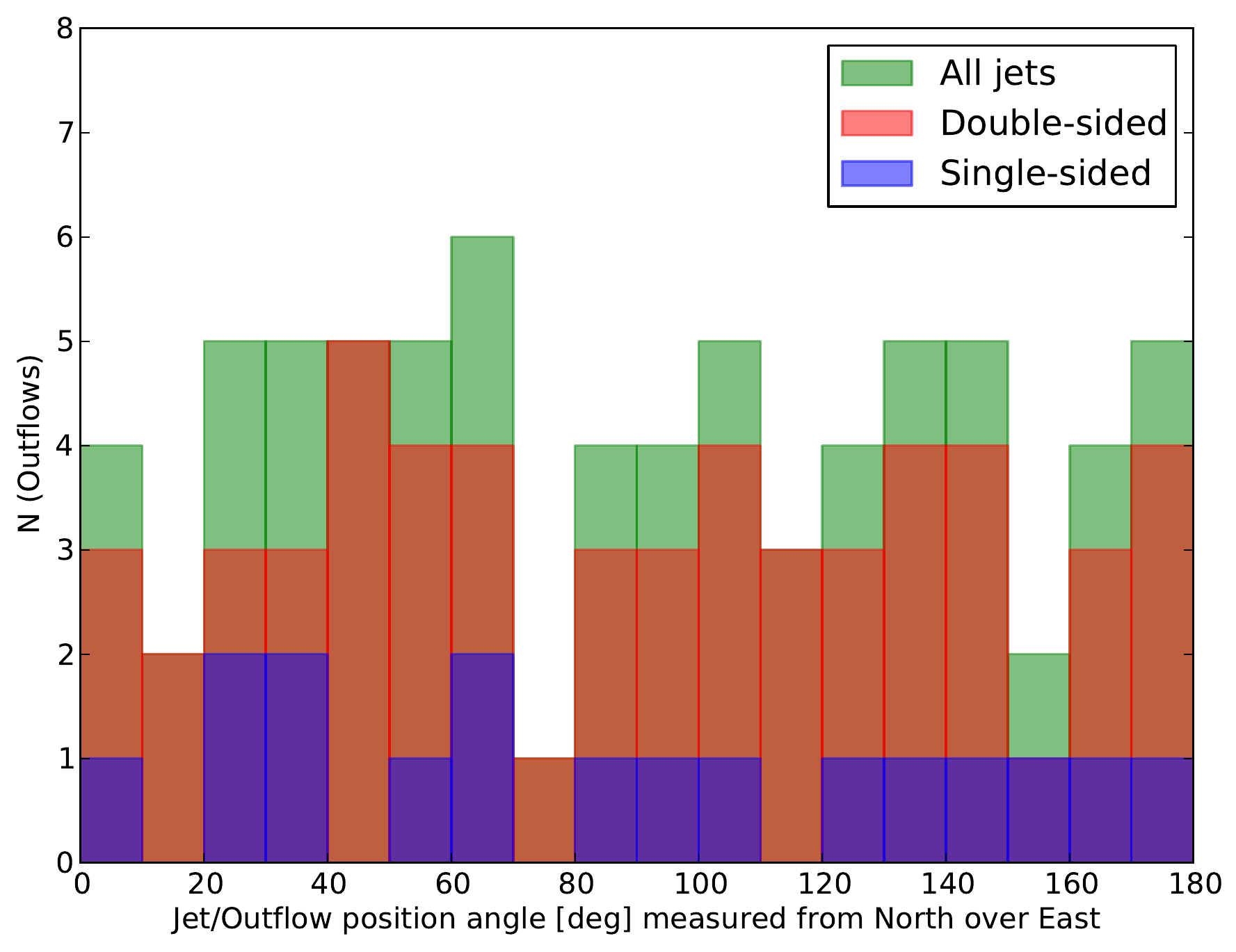} 

\caption{Outflow position angle distribution for all MHOs with a driving source
candidate. The different coloured histograms indicate the single-sided and bipolar,
as well as all outflows. \label{orientationdistribution_figure}}

\end{figure}

We have measured the position angle of all single-sided and bipolar outflows as
the angle between the vector pointing from the source candidate along the
outflow lobe and the North direction (towards East). For bipolar outflows the
position angles of the two lobes are measured separately and are averaged. In a
final step the position angles are taken modulus 180\degr. In
Fig.\,\ref{orientationdistribution_figure} we show the distributions of the
position angles separated for single-sided and bipolar outflows.

The distribution shows no apparent inhomogeneities. We used a 2-sample
Kolmogorov-Smirnov (KS) test to investigate if the position angles are in
agreement with a homogeneous distribution of angles between 0\degr\ and
180\degr. For the bipolar outflows we find a probability $p > 0.9999$ that the
position angle distribution and a homogeneous distribution are drawn from the
same parent sample. For the single-sided outflows we find $p = 0.986$ and for
all outflows combined $p = 0.997$. Furthermore, $p = 0.988$ that the position
angle distribution for bipolar and single-sided outflows is drawn from the same
parental distribution. We also determined the vector representing the mean
position angles (taken as unit vectors) and of all lobes of the bipolar
outflows. This mean angle vector has a length of 0.009 units, hence showing as
well that the position angles are homogeneously distributed.

We also investigate the differences in position angles for the two sides of the
bipolar outflows. Typically the outflows are straight, with a median difference
in the position angles between the two lobes of 5.0\degr. There are only a few
objects (8/57) for which the two sides have a position angle difference of more
than 10\degr.

In total 12 (12\,\%) of the MHOs are part of X-shaped regions (e.g. MHO\,1092,
1093, 1094) where there seem to be two or more outflows originating from the same
source at about 90\degr from each other. \citet{2016ApJ...820L...2L} have shown
that binary outflows seem to be orientated randomly or even preferentially
perpendicular with respect to each other. Hence, all these groups could be
originating from binaries. Another typical sign for a binary source is precession
of the outflow. In total there are only 4 MHOs with clear signs of precession.
Half of them are part of the X-shaped regions.

\subsubsection{Jet/Outflow length}\label{sect_rl}

\begin{figure}
\centering
\includegraphics[angle=0,width=\columnwidth]{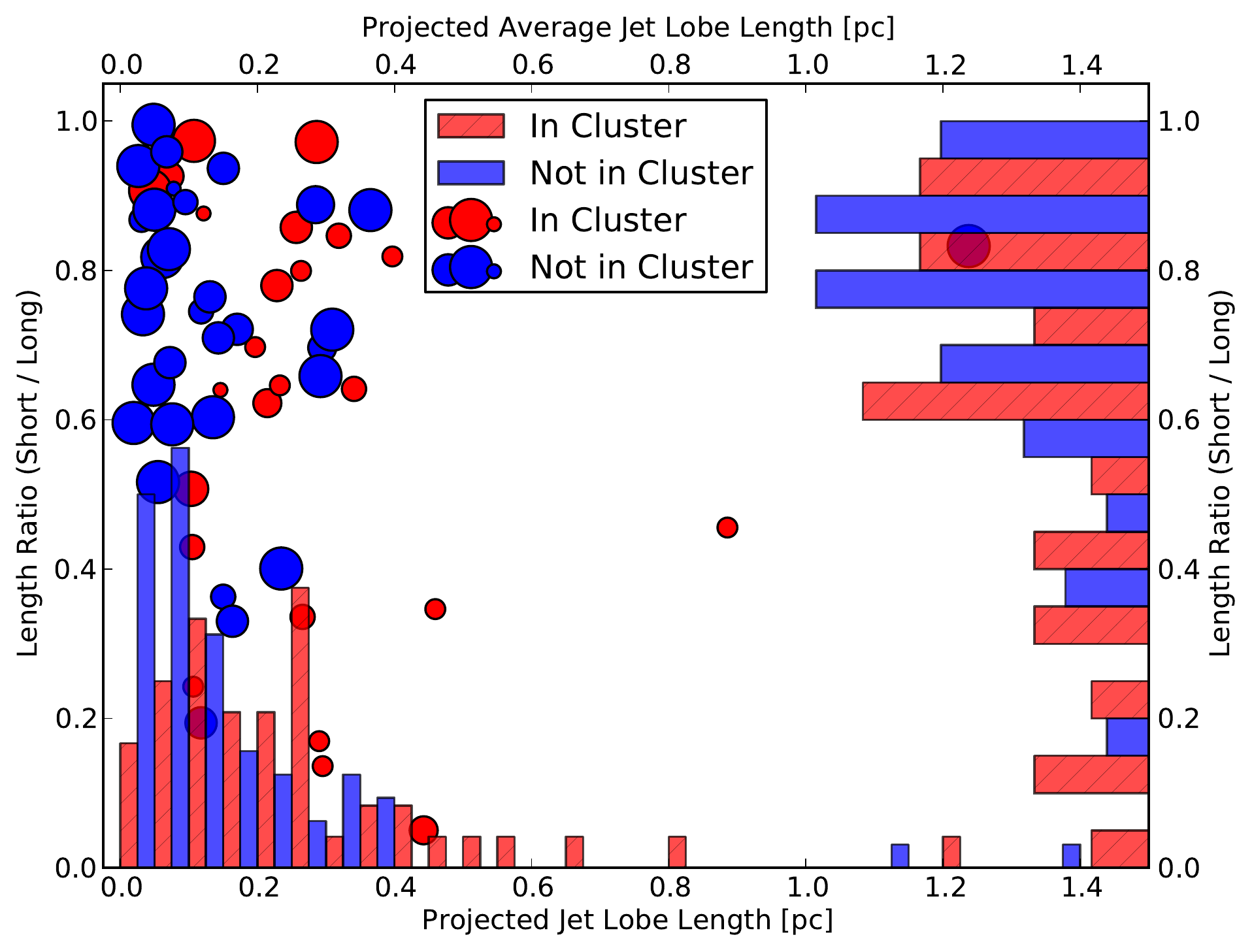} 

\caption{The circles indicate the length ratio (short over long) of the two
lobes of the detected bipolar outflows plotted against the average lobe length
of the MHOs. The circle size is proportional to the source probability. The
histograms show the distributions of the projected lobe lengths and length
ratio. The histograms for sources in clusters and not in clusters are normalised
to the same number of objects. \label{lengthratio_figure}}

\end{figure}

We have measured the lengths of both outflow lobes in all the bipolar objects
identified. In Fig.\,\ref{lengthratio_figure} we show the length ratio ($R_L$),
which we define as the ratio of the length of the shorter lobe and the length of
the longer lobe. Hence, completely symmetrical objects have a length ratio of
one, and all values are between zero and one. As one can see, there is an almost
homogeneous distribution of $R_L$ values between 0.6 and 1.0. The median length
ratio of all bipolar objects is $R_L = 0.72$. Hence, the typical bipolar outflow
is somewhat asymmetric. However, highly asymmetric objects are rare. For example
there are only 12/57\,=\,21\,\% of objects with $R_L < 0.5$, and a large
fraction of these have low probability source candidates.

We find that $p = 0.135$ when we test if the $R_L$ distributions for bipolar
objects in and not in clusters are drawn from the same parent distribution. There
seems to be a slightly higher fraction of asymmetric objects associated with
clusters. If the structure of the environment is responsible for the formation of
bright \htwo\ knots, then a more clumpy interstellar medium (expected near
clusters) would lead to more asymmetric length ratios. However, as evident from
Fig.\,\ref{lengthratio_figure} most of the driving source candidates of the
asymmetric outflows have a low source probability. Hence there is currently no
statistically significant difference in the asymmetry distribution of outflows
from clustered sources and isolated objects.

\subsubsection{Jet/Outflow length distribution}\label{sect_length}

\begin{figure*}
\centering
\includegraphics[angle=0,width=\columnwidth]{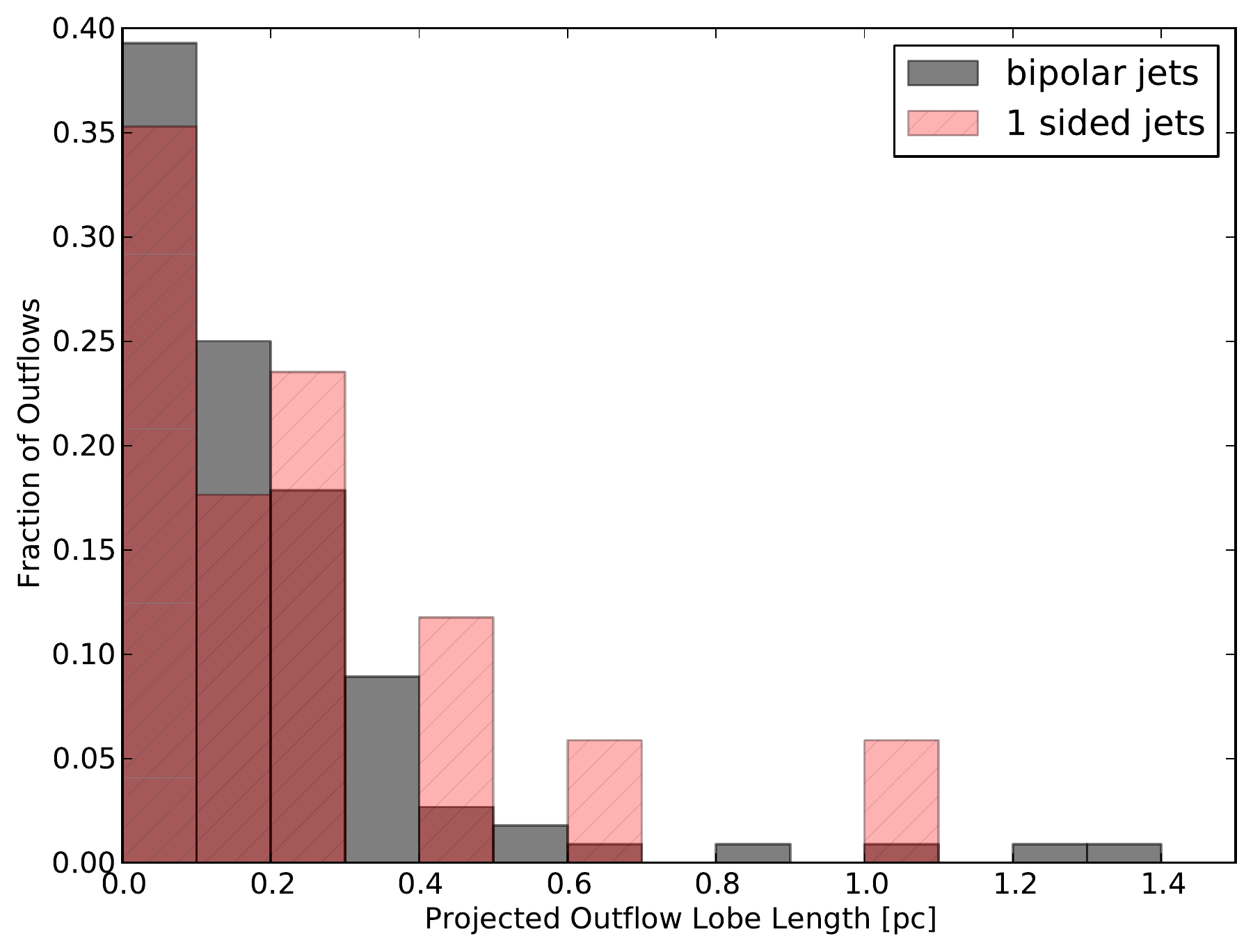} \hfill
\includegraphics[angle=0,width=\columnwidth]{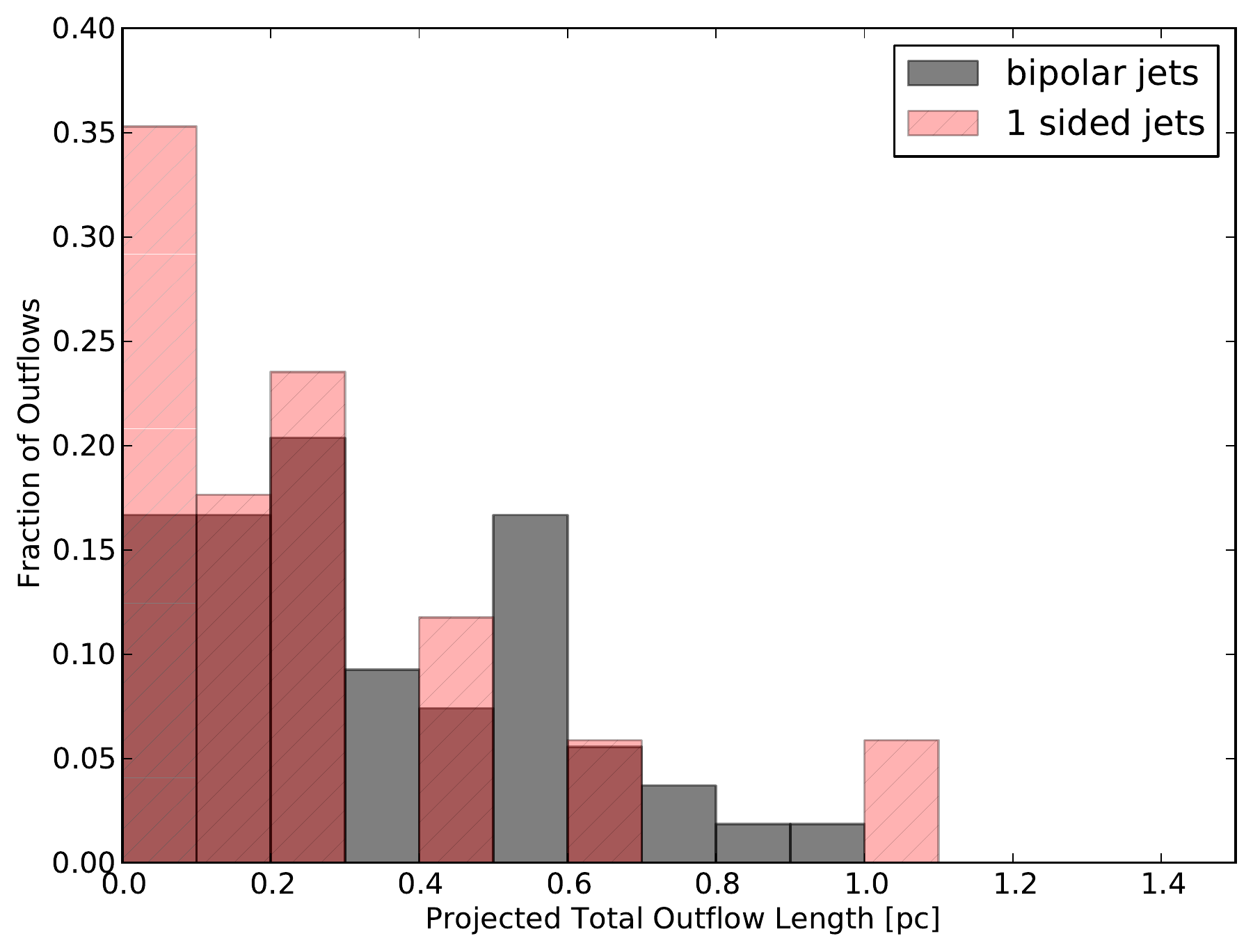} 

\caption{Distribution of outflow lobe (left) and total, end-to-end (right)
lengths for bipolar and single-sided objects. All histograms are normalised to
the same total number of objects. \label{ld_single_bipolar}}

\end{figure*}

In the left panel of Fig.\,\ref{ld_single_bipolar} we show the distribution of
projected lobe lengths for the outflows, separated into bipolar and single-sided
objects. As expected there are more shorter outflows than longer ones. It seems
that the fraction of longer outflow lobes is increased for the single-sided
objects. But small number statistics is an issue and a KS-test is also indecisive.
We find  $p = 0.53$ that both distributions are drawn from the same parent sample.
Note that some of this could be explained by potential systematic
misidentification of driving sources for the single-sided objects.

As most of the literature uses the total lengths of jets and outflows
(end-to-end for bipolars and source-to-end of single-sided objects) to
investigate the length distributions we compare them in the right panel of
Fig.\,\ref{ld_single_bipolar}. As one can see, the total length distribution for
the bipolars differs significantly from the lobe length distribution. There are
far fewer short objects when measuring the total length. The KS-test shows that
the total length distribution of the single-sided and the bipolar outflows are
most likely drawn from different distributions ($p = 0.18$). Hence, mixing the
two types of objects into a common length distribution should be avoided. 
Furthermore, due to the typical asymmetry of the bipolar jets (see
Sect.\,\ref{sect_rl}), the lobe length and total length distributions of the
bipolars are very different ($p = 4.644 \times 10^{-4}$).

We compare the lobe length distribution of the bipolar outflows with sources
associated with clusters and those not. The two distributions are shown as
histograms in Fig.\,\ref{lengthratio_figure}. The fraction of objects with a
short lobe length is much higher for objects not associated with clusters. The
KS-test results in $p = 0.0014$ that the distributions are from the same parent
sample. Hence, the more isolated bipolar outflows seem to be shorter. Indeed the
median lobe length for clustered objects is 0.22\,pc, while for the
non-clustered objects the median is only 0.09\,pc, less than half.  This, to
some extent, can be explained by the larger fraction of highly asymmetric
outflows found to be associated with clusters (see Sect.\,\ref{sect_rl}). Hence,
objects associated with clusters will have one of their lobes much longer than
the average. There is a possibility that some of this is a distance bias, i.e.
apparently isolated objects are more distant, and hence the clusters are not
visible. We cannot rule this out entirely, as there are no individual distances
for all objects and the distance measurement method applied by
\citet{2012MNRAS.425.1380I} will not work in this area due to the lack of
sufficient foreground stars.

We also tested if the lobe length distributions for outflows from younger and
older driving sources are different. The KS test is indecisive leading to $p =
0.55 - 0.65$, depending on whether all lobe lengths or only the bipolar outflows
are considered. The only slight difference between the two groups of lobe
lengths is that most of the longest ($> 0.5$\,pc) lobes are associated with
older driving sources.

\subsubsection{Gaps between \htwo\ knots}

\begin{figure}
\centering
\includegraphics[angle=0,width=\columnwidth]{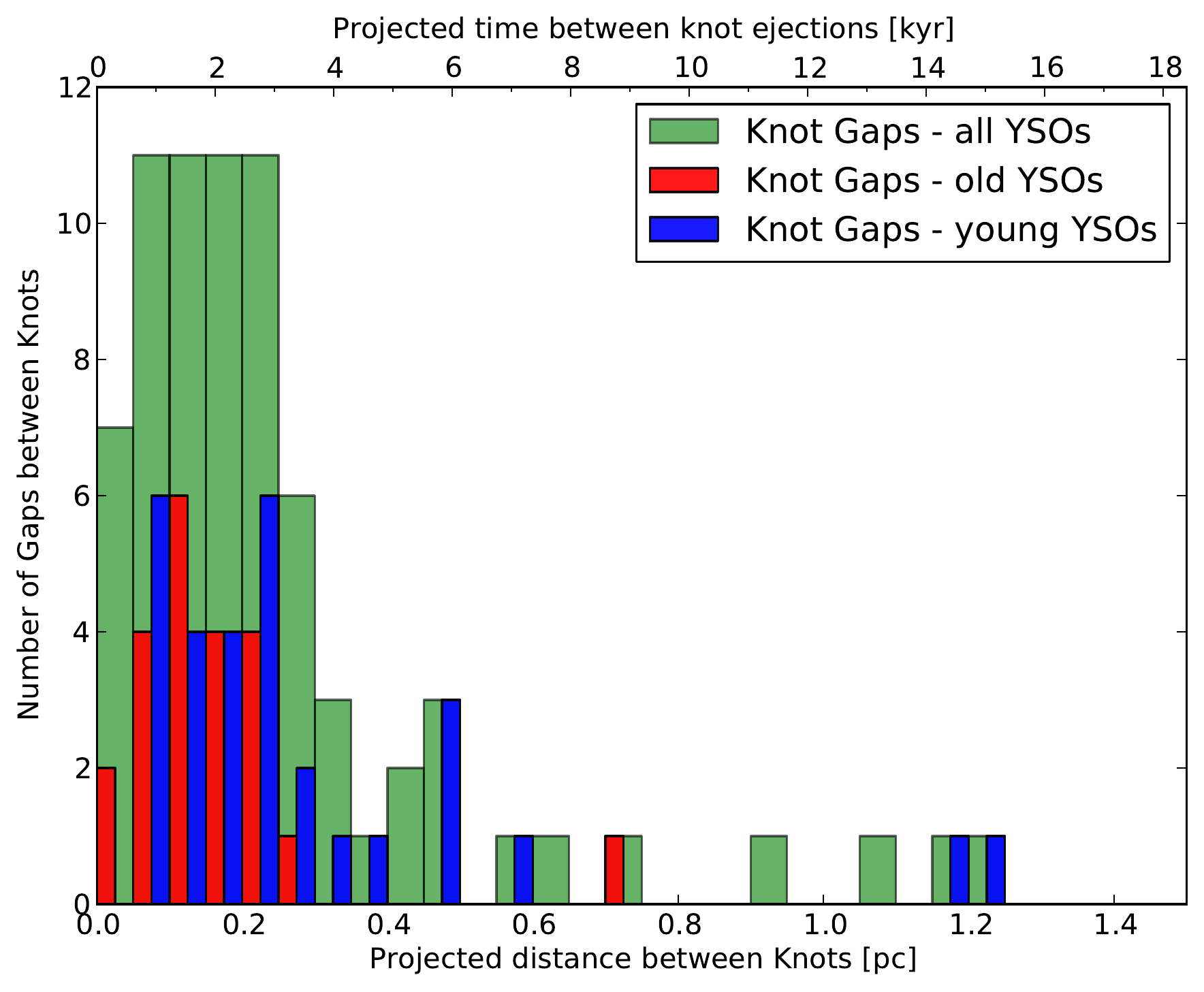} 

\caption{Distribution of gaps between individual outflow knots. Younger and older
YSOs are separated using alpha=-0.5. \label{gapdistribution_figure}}

\end{figure}

In Fig.\,\ref{gapdistribution_figure} we show the distribution of gaps between
the main (isolated) \htwo\ emission knots in the jets and outflows. These larger
knots represent either major mass ejection/accretion events or the position of
denser regions in the environment. The median gap is about 0.19\,pc, and the
mean is 0.25\,pc, which converts to about 2\,--3\,kyr between the knots at the
assumed distance of 2\,kpc and a projected transversal speed of 80\,km/s 
(assumed to be constant). There is a range of the most common knot gaps between
0.05\,pc and 0.25\,pc, corresponding to 1\,--\,3\,kyr between ejections. These
numbers are similar to what has been found for outflows in Serpens and Aquila by
\citet{2012MNRAS.425.1380I}, who showed that the typical gaps between large
\htwo\ knots correspond to 1\,--\,2\,kyr.

As seen in Sect.\,\ref{sect_length}, the dynamical timescales of the outflow
lobes, estimated from their typical total lengths, amount to values that range
from the above determined few kyr to about 10\,kyr for the parsec scale
outflows. Hence, if the main knots in outflows are generated by major mass
ejection events, then they should occur typically every few kyr for the
jet-driving young stellar objects. Given the estimates for the FU-Ori occurrence
rates of 5\,--\,50\,kyr \citep{2013MNRAS.430.2910S}, it is highly unlikely that
FU-Ori type events are the cause of the main \htwo\ knots. Similarly, EX-Ors are
also not suitable candidates for causing these knots, as these bursts repeat
semi-regularly every one to ten years.  Recent NIR monitoring of large parts of
the Galactic Plane (in UKIDSS GPS, \citet{2014MNRAS.439.1829C}; VISTA VVV,
\citet{2016arXiv160206267C}) reveals a large number of variables with
characteristics in-between the FU-Ori and EX-Or classifications. Dubbed MNors by
\citet{2016arXiv160206269C}, these apparently more common eruptions might be the
cause of the main \htwo\ knots in outflows. Further statistics are required
however, to establish if the occurrence rate of the MNor eruptions correspond to
the typical time gaps between the outflow knots.

We separated the gaps in the outflows based on the slope of the spectral energy
distribution of the driving source. We use $\alpha$\,=\,-0.5 as the separator of
young and old objects. The two distributions for the gaps between knots are also
shown in Fig.\,\ref{gapdistribution_figure}, where each main bin is separated
into the two bins for younger and older sources. While the basic distributions
seem to look the same, they are in fact different ($p = 0.14$). Both the mean
and median gap sizes decrease by 70\,\% from the younger to the older driving
sources. Of the ten largest gaps between knots, only one of them is associated
with an older driving source. 

Despite the low $p$-value it is not clear if this difference is caused by small
number statistics or indeed real. In a future paper we will study the outflows
in the Cygnus\,X region, where we have identified between four and five times
as many objects as in Cassiopeia and Auriga. One would expect that younger
sources have a higher frequency of accretion bursts and hence mass ejection
events (e.g. \citet{Vorobyov2006}). Thus, the gaps between knots should be, on
average, smaller in outflows from younger driving sources if the typical
ejection speeds are comparable. Even if the density structure near the driving
sources governs the \htwo\ knot formation, one would expect a more clumpy
medium near younger sources and hence smaller gaps between knots.

One can combine the results from Sect.\,\ref{sect_length} (longer outflows
originate more commonly from older sources) and this section (there are smaller
gaps between knots in outflows from older objects). This leads to the conclusion
that on average, there are more knots per outflow lobe length in the outflows
from older sources. Thus, even considering potentially variable speeds for the
ejection of material, the frequency of the formation of larger \htwo\ knots is
higher for older driving sources. Note that this is based on small number
statistics and will be revisited in our future paper discussing the outflows
identified in Cygnus\,X. Furthermore, there is a possibility that some of the
knots from younger sources are not detected due to higher extinction levels. 

\begin{figure}
\centering
\includegraphics[angle=0,width=\columnwidth]{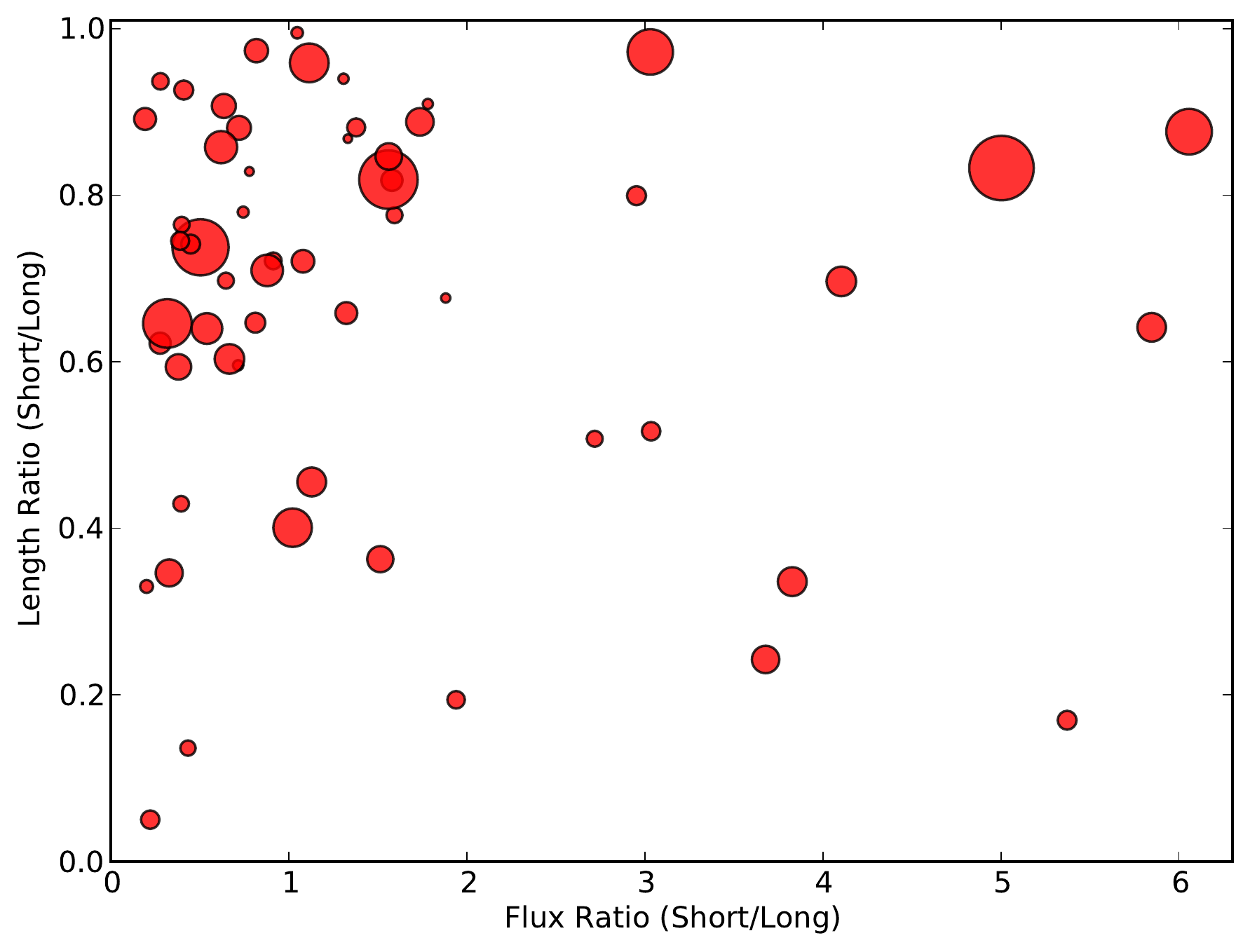} 

\caption{Jet length ratio plotted against the flux ratio of the lobes for all
bipolar jets. The size of the circles indicates the total flux of the outflow.
\label{lengthfluxratio_figure}}

\end{figure}

\begin{figure*}
\centering
\includegraphics[angle=0,width=\columnwidth]{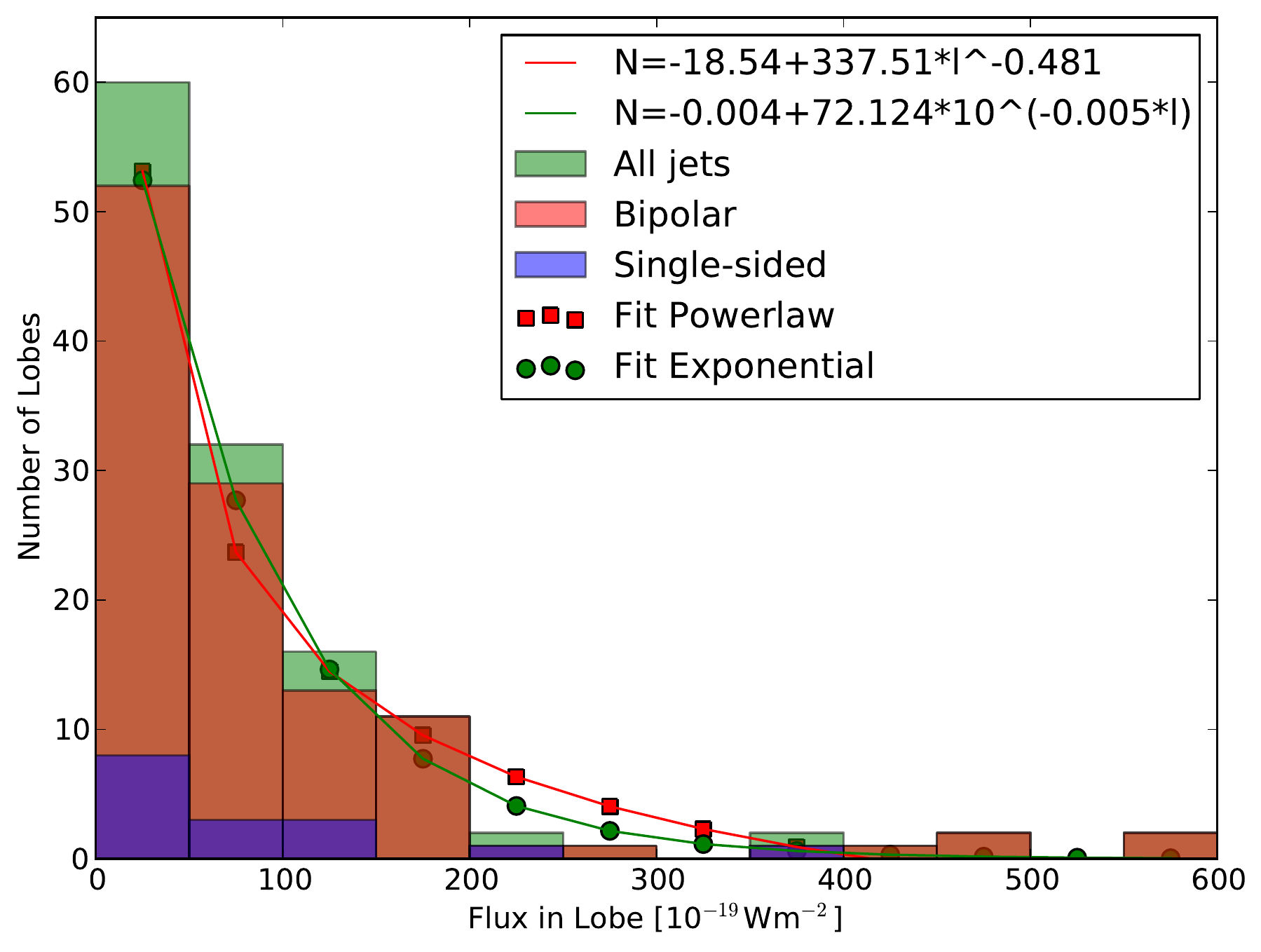} \hfill
\includegraphics[angle=0,width=\columnwidth]{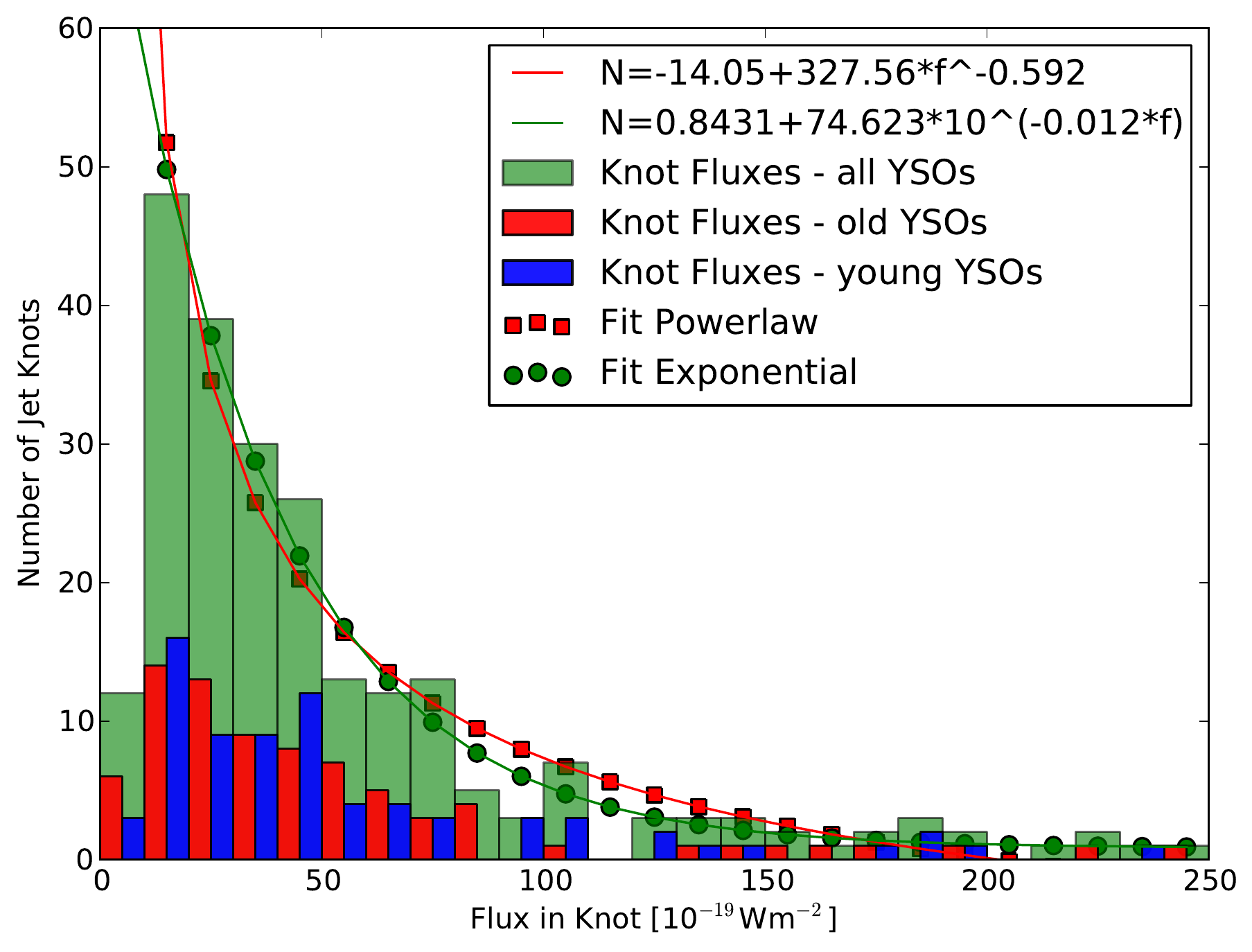} 

\caption{Flux distribution of the outflow lobes (left) and the individual knots
(right). Some objects have fluxes higher than the largest values along the
x-axis in the graphs and are not shown. \label{fluxlobe_distribution_figure}}

\end{figure*}

\subsubsection{Jet/Outflow flux distribution}

Figure\,\ref{lengthfluxratio_figure} shows that there is no correlation between
flux ratio and length ratio of the two lobes in bipolar outflows. Both ratios
are estimated as the values (length or flux) of the short side divided by the
value of the long side, hence all length ratios are less than one. About half of
the bipolar objects have a flux ratio less than one and half a flux ratio larger
than one. This percentage also does not depend on the flux itself. Essentially,
the brightness ratio of the two lobes in bipolar jets/outflows is statistically
independent of the length ratio of the same two lobes. 

In Fig.\,\ref{fluxlobe_distribution_figure} we show the flux distributions of
all \htwo\ knots (right panel) as well as the flux distributions of all outflow
lobes (left panel). Generally the flux distributions are better fit by
exponentials than power laws, due to a number of bright 'outliers'. They are too
frequent for the entire distribution to be power laws and are caused by some of
the very long outflows in our sample. Generally the $rms$ of the power law fits
are about twice as high as for the exponential fits of the distributions.
Essentially distributions start to deviate from power laws at fluxes above $15
\times 10^{-18}$\,W\,m$^{-2}$ for the individual \htwo\ knots and $30 \times
10^{-18}$\,W\,m$^{-2}$ for the fluxes in the outflow lobes. When fit by power
laws, the lobe flux distribution follows $N \propto F^{-0.5}$ while the flux
distribution of the individual knots in the outflows is a bit steeper with $N
\propto F^{-0.6}$. These numbers do not change if only those objects below the
above quoted flux limits are used for the fits. However, in these cases the
exponential fits are still resulting in a lower $rms$. Generally the power law
exponents are in agreement with the values for the luminosity distributions of
outflows in Serpens and Aquila by \citet{2012MNRAS.425.1380I} who found $N
\propto F^{-0.5 ... -0.7}$. Note that these are measured as a mix of bipolar and
single-sided outflows.

The measured fluxes of the outflows can be converted into luminosities with our
assumed distance for all objects of 2\,kpc. In this case, a flux of $5 \times
10^{-18}$\,W\,m$^{-2}$ (corresponding to the median lobe flux in our sample)
converts to about $0.6 \times 10^{-3}$\,L$_\odot$ in the 1\,--\,0\,S(1) line of
\htwo. This median luminosity is at the faint end of the luminosity
distribution of outflows investigated along the Galactic plane in Serpens and
Aquila by \citet{2012MNRAS.425.1380I}. This is caused by the fact that the
typical distances for the objects in this study were 3\,--\,5\,kpc (compared to
our 2\,kpc) while the median flux of the two samples is almost identical. Thus,
the MHO sample investigated here includes a number of intrinsically fainter
objects that fell below the detection limit in the study by
\citet{2012MNRAS.425.1380I}. However, the number of MHOs with a brightness
above $10^{-3}$\,L$_\odot$, normalised to the distance-corrected projected
survey area is comparable in both studies. In Cassiopeia and Auriga this number
is only about 30\,\% lower than in Serpens and Aquila. 

We investigated if the evolutionary stage ($\alpha$) of the driving sources
influences the flux distributions. However, as can be seen in the right panel of
Fig.\,\ref{fluxlobe_distribution_figure} there are no differences in the flux
distribution of \htwo\ knots of outflows from younger and older driving sources
($p = 0.94$). Similarly the flux distribution for the outflow lobes of younger
and older driving sources are similar with $p=0.93$.

The right-hand side histogram in Fig.\,\ref{fluxlobe_distribution_figure} also
indicates that our completeness limit for the detection of outflow knots is of
the order of $10^{-18}$\,W\,m$^{-2}$. This compares well with the estimated
5\,$\sigma$ surface brightness detection limit in UWISH2 of $4.1 \times
10^{-19}$\,W\,m$^{-2}$\,arcsec$^{-2}$ \citep{2015MNRAS.454.2586F}, since most of
the knots are extended over more than a few square arcseconds.

\section{Conclusions}

We have analysed the population of jets and outflows from young stars in a 35.5
square degree region along the Galactic Plane in Cassiopeia and Auriga utilising
\htwo\ imaging data from the UWISH2 survey \citep{Froebrich2011} and the
catalogue of potential jet/outflow features from \citet{2015MNRAS.454.2586F}. In
the investigated area we have identified 98 Molecular Hydrogen emission-line
Objects, i.e. potential jets and outflows from young stars, all of which are new
discoveries. When scaled up to the entire UWISH2 survey, we thus expect a total
sample of about 1500 MHOs for analysis. 

The detected MHOs are classified as bipolar outflows (60\,\%), single-sided
outflows (20\,\%) and individual or small groups of \htwo\ knots (20\,\%). 
Comparing the number of identified MHOs to the automatically generated and
classified catalogue of UWISH2-detected \htwo\ features, we can conclude that
the catalogue from \citet{2015MNRAS.454.2586F} contains much less than 10\,\% of
false positives amongst the \htwo\ features classified as jets and outflows from
young stars. Most of these false positives can be attributed to fluorescently
excited edges of molecular clouds. The entire \htwo\ feature catalogue contains
only 2\,\% of false positives, i.e. features that are not caused by emission
from \htwo. These are mostly variable stars and image artefacts. Finally, the
catalogue of jet/outflow features in \citet{2015MNRAS.454.2586F} is complete at
the 95\,\% level for objects above the UWISH2 detection limit. Only a small
number of low surface brightness features detectable in the UWISH2 data have not
been included in the automatically generated catalogue.

We could identify driving source candidates for about 75\,\% of the MHOs and
about 40\,\% of these objects are associated with groups or clusters of stars,
while the remaining 60\,\% seem to be more isolated, indicating that clustered
star formation could inhibit the formation of detectable larger outflows. We
have identified 15 new star cluster candidates near the MHOs in the survey
area. Of the WISE detected driving source candidates, about 20\,\% have
positive slopes of their SED, i.e. are protostellar source candidates, while
the remaining 80\,\% are most likely CTTSs. This is also supported by their NIR
colours. 70\,\% of the driving sources with multiple NIR K-band detections
(UKIDSS GPS and 2MASS) show a variability of more than 0.1\,mag and about
20\,\% of the sources vary by more than 0.5\,mag over a timescale of several
years. 

The position angles of the identified outflows have a 99.99\,\% probability to
be drawn from a homogeneous distribution. For the typical bipolar outflow the
two lobes have position angles within 5\degr. About 10\,\% of the MHOs form
X-shaped outflows which could be originating from binary sources
\citep{2016ApJ...820L...2L}. 

The length ratio (short over long) of the lobes of the bipolar outflows have
typical values between 0.6 and 1.0, with a median of 0.72. Only about 20\,\% of
objects are highly asymmetric with length ratios of less than 0.5. The flux
ratios (short over long) typically show a much wider spread (between 0.2 and
5.0) and are not correlated with the length ratio.

We measured the length of all outflow lobes and investigated their
distributions. We find that the length measurements typically used in the
literature (end-to-end for bipolar outflows and source-to-end for single-sided
flows) should be avoided. This is due to the typical asymmetry of the bipolar
outflows. It  causes the lobe length distributions and total length
distributions to be different with a probability of 99.95\,\%. There is no
apparent difference in the lobe length distributions of outflows from younger
and older driving source candidates.

The dynamical timescales of the outflows are up to 10\,kyr, while the typical
timescales associated with the gaps of large \htwo\ knots in the lobes
correspond to 1\,--\,3\,kyr. This indicates that neither FU-Ori or EX-Ori style
outbursts are likely to be responsible for the formation of the larger \htwo\
knots seen in the typical outflows. Potentially the population of recently
identified NIR eruptive variables, or MNors, with properties in-between FU-Ori
and EX-Ori objects, could be the cause for the \htwo\ knot formation, if their
as-yet unknown occurrence rate has the correct timescale.

The flux distributions of \htwo\ knots and outflow lobes is generally better fit
by an exponential than by a power law. This is caused by a increased number of
bright knots and lobes (above $30 \times 10^{-18}$\,W\,m$^{-2}$) compared to the
expectation for a power law distribution. There are no differences in the flux
distribtions for outflows from younger or older driving sources. The
completeness limit for the detection of \htwo\ knots in the outflows is
estimated as $10^{-18}$\,W\,m$^{-2}$. The number of \htwo\ bright ($>
10^{-3}$\,L$_\odot$ in the 1\,--\,0\,S(1) line) outflows per square parsec in
Aquila is comparable to investigations in the inner Galactic Plane.

In a future paper we will investigate these trends for a much larger sample of
MHOs detected in the Cygnus\,X region where we expect a four- to five-fold
increase in the number of MHOs. This will also be used to investigate if any of
the trends are caused by environmental effects.

\section*{acknowledgements}

S.V.\,Makin acknowledges an SFTC scholarship (1482158). The United Kingdom
Infra-Red Telescope is operated by the Joint Astronomy Centre on behalf of the
Science and Technology Facilities Council of the U.K. The data reported here
were obtained as part of the UKIRT Service Program. 

\bibliographystyle{mn2e}
\bibliography{biblio}

\clearpage
\newpage

\onecolumn

\begin{appendix}
\section{MHO data table}

\renewcommand{\tabcolsep}{2pt}
\setlength\LTcapwidth{\textheight}

\begin{landscape}
\begin{longtable}{|c|c|c|r|r|r|c|c|c|c|r|r|r|r|r|r|r|c|c|c|c|c|}

\caption{Summary table of the MHO properties and their respective candidate
driving sources. The table contains the following columns: i) the MHO number;
ii) Right Ascension of the object (J2000) in degrees; iii) Declination of the
object (J2000) in degrees; iv) Length of the outflow lobe(s) in degrees; v)
Position angle of the outflow lobe(s) in degrees from North over East; vi)
{Integrated flux of the outflow lobe(s) or knots;} vii) Type of outflow (B -
bipolar, S - single-sided, K - single or group of knots without apparent source
candidate); viii) Is the MHO associated with cluster or group of stars? ix)
Source candidate identification; x) Likelihood P$_S$ that source candidate is
the real driving source in percent;  xi)\,--\,xiii) Near-infrared JHK magnitudes
of the source candidate; xiv)\,--\,xvii) Mid-infrared WISE magnitudes of the
source candidate; xviii)\,--\,xxii) Is there a detection of the source candidate
in the following surveys/catalogues: G -- UKIDSS GPS, 2 -- 2MASS, W -- WISE, A
-- AKARI, I -- IRAS? $^{(1)}$ There are two values for all bipolar MHOs, one for
each outflow lobe; $^{(2)}$ The unit of the fluxes is 10$^{-19}$\,W\,m$^{-2}$;
$^{(3)}$ The first number refers to the magnitude in UKIDSS GPS, the second
number to 2MASS. Note that same 2MASS detections might be blends of several
sources. There are 4 objects with 2 GPS K-band epochs, we only list the first
one in these cases as there is no variability above the 0.1\,mag level between
the two epochs in all cases.\label{datatable}}

\label{mainresults}\\

\hline 
\multicolumn{1}{|c|}{\textbf{MHO}} &
\multicolumn{2}{c|}{\textbf{Position (J2000)}} & 
\multicolumn{1}{c|}{\textbf{Length$^{(1)}$}} & \multicolumn{1}{c|}{\textbf{PA$^{(1)}$}} &
\multicolumn{1}{c|}{\textbf{Flux$^{(1,2)}$}} & \multicolumn{1}{c|}{\textbf{Type}} &
\multicolumn{1}{c|}{\textbf{Cl.}} & \multicolumn{1}{c|}{\textbf{Source}} &
\multicolumn{1}{c|}{\textbf{P$_{\rm S}$}} &
\multicolumn{1}{c|}{\textbf{J$^{(3)}$}} & \multicolumn{1}{c|}{\textbf{H$^{(3)}$}}
& \multicolumn{1}{c|}{\textbf{K$^{(3)}$}} & \multicolumn{1}{c|}{\textbf{W1}} &
\multicolumn{1}{c|}{\textbf{W2}} & \multicolumn{1}{c|}{\textbf{W3}} &
\multicolumn{1}{c|}{\textbf{W4}} & \multicolumn{1}{c|}{\textbf{G}} &
\multicolumn{1}{c|}{\textbf{2}} & \multicolumn{1}{c|}{\textbf{W}} &
\multicolumn{1}{c|}{\textbf{A}} & \multicolumn{1}{c|}{\textbf{I}} \\

\multicolumn{1}{|c|}{\textbf{}} & \multicolumn{2}{c|}{\textbf{[deg]}} & 
\multicolumn{1}{c|}{\textbf{[deg]}} &\multicolumn{1}{c|}{\textbf{[deg]}} &
\multicolumn{1}{c|}{\textbf{}} & \multicolumn{1}{c|}{\textbf{B/S/K}} &
\multicolumn{1}{c|}{\textbf{Y/N}} & \multicolumn{1}{c|}{\textbf{ID}} &
\multicolumn{1}{c|}{\textbf{[\%]}} & \multicolumn{3}{c|}{\textbf{[mag]}} &
\multicolumn{4}{c|}{\textbf{[mag]}} & \multicolumn{5}{c|}{\textbf{Y/N}} \\
\hline 

\endfirsthead

\multicolumn{22}{c}%
{{\bfseries \tablename\ \thetable{} -- continued from previous page}} \\

\hline \multicolumn{1}{|c|}{\textbf{MHO}} &
\multicolumn{2}{c|}{\textbf{Position (J2000)}} & 
\multicolumn{1}{c|}{\textbf{Length$^{(1)}$}} & \multicolumn{1}{c|}{\textbf{PA$^{(1)}$}} &
\multicolumn{1}{c|}{\textbf{Flux$^{(1,2)}$}} & \multicolumn{1}{c|}{\textbf{Type}} &
\multicolumn{1}{c|}{\textbf{Cl.}} & \multicolumn{1}{c|}{\textbf{Source}} &
\multicolumn{1}{c|}{\textbf{P$_{\rm S}$}} &
\multicolumn{1}{c|}{\textbf{J$^{(3)}$}} & \multicolumn{1}{c|}{\textbf{H$^{(3)}$}}
& \multicolumn{1}{c|}{\textbf{K$^{(3)}$}} & \multicolumn{1}{c|}{\textbf{W1}} &
\multicolumn{1}{c|}{\textbf{W2}} & \multicolumn{1}{c|}{\textbf{W3}} &
\multicolumn{1}{c|}{\textbf{W4}} & \multicolumn{1}{c|}{\textbf{G}} &
\multicolumn{1}{c|}{\textbf{2}} & \multicolumn{1}{c|}{\textbf{W}} &
\multicolumn{1}{c|}{\textbf{A}} & \multicolumn{1}{c|}{\textbf{I}} \\

\multicolumn{1}{|c|}{\textbf{}} & \multicolumn{2}{c|}{\textbf{[deg]}} & 
\multicolumn{1}{c|}{\textbf{[deg]}} &\multicolumn{1}{c|}{\textbf{[deg]}} &
\multicolumn{1}{c|}{\textbf{}} & \multicolumn{1}{c|}{\textbf{B/S/K}} &
\multicolumn{1}{c|}{\textbf{Y/N}} & \multicolumn{1}{c|}{\textbf{ID}} &
\multicolumn{1}{c|}{\textbf{[\%]}} & \multicolumn{3}{c|}{\textbf{[mag]}} &
\multicolumn{4}{c|}{\textbf{[mag]}} & \multicolumn{5}{c|}{\textbf{Y/N}} \\
\hline 

\endhead

\hline \multicolumn{22}{|r|}{{Continued on next page}} \\ \hline
\endfoot

\hline
\endlastfoot

$\begin{array}[t]{c}{\rm MHO}\\1064\end{array}$ & 60.23841   & 53.356366  & $\begin{array}[t]{r}0.00417\\ 0.00445\end{array}$   & $\begin{array}[t]{r}338.1\\166.3\end{array}$ & $\begin{array}[t]{r}  16\\   57\end{array}$ & B & N & GPS439233255353      & 50 & $\begin{array}[t]{r}    -\\     -\end{array}$ & $\begin{array}[t]{r}18.46\\    -\end{array}$ & $\begin{array}[t]{r}16.95\\     -\end{array}$ & 16.268 & 14.680 & 11.943 &  8.574 & Y & N & Y & N & N \\
$\begin{array}[t]{c}{\rm MHO}\\1065\end{array}$ & 60.39731   & 51.826319  & $\begin{array}[t]{r}0.00483\\ 0.00291\end{array}$   & $\begin{array}[t]{r}318.3\\137.4\end{array}$ & $\begin{array}[t]{r} 140\\   93\end{array}$ & B & N & GPS439239324153      & 90 & $\begin{array}[t]{r}19.48\\     -\end{array}$ & $\begin{array}[t]{r}17.21\\    -\end{array}$ & $\begin{array}[t]{r}15.55\\     -\end{array}$ & 13.899 & 11.382 &  8.249 &  4.871 & Y & N & Y & Y & Y \\
$\begin{array}[t]{c}{\rm MHO}\\1066\end{array}$ & 60.06662   & 51.504271  & $\begin{array}[t]{r}0.00151\\ 0.00133\end{array}$   & $\begin{array}[t]{r}  5.6\\188.7\end{array}$ & $\begin{array}[t]{r}  37\\   51\end{array}$ & B & N & noname               & 90 & $\begin{array}[t]{r}    -\\     -\end{array}$ & $\begin{array}[t]{r}    -\\    -\end{array}$ & $\begin{array}[t]{r}    -\\     -\end{array}$ & 14.600 & 13.658 & 12.568 &  9.142 & N & N & Y & Y & N \\
$\begin{array}[t]{c}{\rm MHO}\\1067\end{array}$ & 60.86852   & 51.863113  & $\begin{array}[t]{r}0.00960\\ 0.00385\end{array}$   & $\begin{array}[t]{r} 70.1\\250.8\end{array}$ & $\begin{array}[t]{r} 195\\  199\end{array}$ & B & N & GPS439239421916      & 90 & $\begin{array}[t]{r}18.59\\ 17.68\end{array}$ & $\begin{array}[t]{r}15.37\\14.82\end{array}$ & $\begin{array}[t]{r}12.81\\ 12.82\end{array}$ & 10.485 &  8.559 &  5.725 &  2.995 & Y & Y & Y & Y & Y \\
$\begin{array}[t]{c}{\rm MHO}\\1068\end{array}$ & 60.83425   & 51.485619  & $\begin{array}[t]{r}0.01141\\ 0.00383\end{array}$   & $\begin{array}[t]{r} 25.8\\208.0\end{array}$ & $\begin{array}[t]{r}  46\\  176\end{array}$ & B & Y & GPS439240314623      & 30 & $\begin{array}[t]{r}    -\\     -\end{array}$ & $\begin{array}[t]{r}18.51\\    -\end{array}$ & $\begin{array}[t]{r}16.36\\     -\end{array}$ & 13.595 & 11.760 &  9.281 &  4.845 & Y & N & Y & Y & N \\
$\begin{array}[t]{c}{\rm MHO}\\1069\end{array}$ & 60.76866   & 51.369731  & $\begin{array}[t]{r}      -\\       -\end{array}$   & $\begin{array}[t]{r}    -\\    -\end{array}$ & $\begin{array}[t]{r}  47\\    -\end{array}$ & K & Y & unknown              &$-$ & $\begin{array}[t]{r}    -\\     -\end{array}$ & $\begin{array}[t]{r}    -\\    -\end{array}$ & $\begin{array}[t]{r}    -\\     -\end{array}$ &    $-$ &    $-$ &    $-$ &    $-$ & N & N & N & N & N \\
$\begin{array}[t]{c}{\rm MHO}\\1070\end{array}$ & 61.01052   & 51.378462  & $\begin{array}[t]{r}0.01250\\ 0.01024\end{array}$   & $\begin{array}[t]{r} 74.1\\245.0\end{array}$ & $\begin{array}[t]{r} 354\\  553\end{array}$ & B & Y & GPS439241003473      & 20 & $\begin{array}[t]{r}17.92\\ 16.86\end{array}$ & $\begin{array}[t]{r}15.28\\15.26\end{array}$ & $\begin{array}[t]{r}13.35\\ 13.34\end{array}$ & 11.209 &  9.418 &  6.419 &  2.522 & Y & Y & Y & Y & N \\
$\begin{array}[t]{c}{\rm MHO}\\1071\end{array}$ & 62.01052   & 52.378462  & $\begin{array}[t]{r}0.01955\\ 0.00677\end{array}$   & $\begin{array}[t]{r}354.8\\167.3\end{array}$ & $\begin{array}[t]{r} 147\\   48\end{array}$ & B & Y & GPS439241003473      & 20 & $\begin{array}[t]{r}17.92\\ 16.86\end{array}$ & $\begin{array}[t]{r}15.28\\15.26\end{array}$ & $\begin{array}[t]{r}13.35\\ 13.34\end{array}$ & 11.209 &  9.418 &  6.419 &  2.522 & Y & Y & Y & Y & N \\
$\begin{array}[t]{c}{\rm MHO}\\1072\end{array}$ & 62.01052   & 52.378462  & $\begin{array}[t]{r}0.01333\\       -\end{array}$   & $\begin{array}[t]{r}324.5\\    -\end{array}$ & $\begin{array}[t]{r} 229\\    -\end{array}$ & S & Y & GPS439241003473      & 20 & $\begin{array}[t]{r}17.92\\ 16.86\end{array}$ & $\begin{array}[t]{r}15.28\\15.26\end{array}$ & $\begin{array}[t]{r}13.35\\ 13.34\end{array}$ & 11.209 &  9.418 &  6.419 &  2.522 & Y & Y & Y & Y & N \\
$\begin{array}[t]{c}{\rm MHO}\\1073\end{array}$ & 61.08369   & 51.391385  & $\begin{array}[t]{r}0.00989\\ 0.00837\end{array}$   & $\begin{array}[t]{r} 16.3\\194.7\end{array}$ & $\begin{array}[t]{r}  73\\  114\end{array}$ & B & Y & J040420.08+512328.9  & 30 & $\begin{array}[t]{r}    -\\     -\end{array}$ & $\begin{array}[t]{r}    -\\    -\end{array}$ & $\begin{array}[t]{r}    -\\     -\end{array}$ & 16.117 & 13.527 & 10.641 &  7.838 & N & N & Y & N & N \\
$\begin{array}[t]{c}{\rm MHO}\\1074\end{array}$ & 61.09325   & 51.396746  & $\begin{array}[t]{r}      -\\       -\end{array}$   & $\begin{array}[t]{r}    -\\    -\end{array}$ & $\begin{array}[t]{r} 140\\    -\end{array}$ & K & Y & unknown              &$-$ & $\begin{array}[t]{r}    -\\     -\end{array}$ & $\begin{array}[t]{r}    -\\    -\end{array}$ & $\begin{array}[t]{r}    -\\     -\end{array}$ &    $-$ &    $-$ &    $-$ &    $-$ & N & N & N & N & N \\
$\begin{array}[t]{c}{\rm MHO}\\1075\end{array}$ & 61.07890   & 51.376090  & $\begin{array}[t]{r}0.00119\\ 0.00131\end{array}$   & $\begin{array}[t]{r} 37.2\\220.4\end{array}$ & $\begin{array}[t]{r}  60\\   94\end{array}$ & B & Y & J040418.93+512233.9  & 90 & $\begin{array}[t]{r}    -\\     -\end{array}$ & $\begin{array}[t]{r}    -\\    -\end{array}$ & $\begin{array}[t]{r}    -\\     -\end{array}$ & 14.306 & 13.644 &  8.544 &  6.434 & N & N & Y & N & N \\
$\begin{array}[t]{c}{\rm MHO}\\1076\end{array}$ & 61.08764   & 51.374903  & $\begin{array}[t]{r}      -\\       -\end{array}$   & $\begin{array}[t]{r}    -\\    -\end{array}$ & $\begin{array}[t]{r}  25\\    -\end{array}$ & K & Y & unknown              &$-$ & $\begin{array}[t]{r}    -\\     -\end{array}$ & $\begin{array}[t]{r}    -\\    -\end{array}$ & $\begin{array}[t]{r}    -\\     -\end{array}$ &    $-$ &    $-$ &    $-$ &    $-$ & N & N & N & N & N \\
$\begin{array}[t]{c}{\rm MHO}\\1077\end{array}$ & 61.48638   & 51.485181  & $\begin{array}[t]{r}      -\\       -\end{array}$   & $\begin{array}[t]{r}    -\\    -\end{array}$ & $\begin{array}[t]{r}  28\\    -\end{array}$ & K & N & unknown              &$-$ & $\begin{array}[t]{r}    -\\     -\end{array}$ & $\begin{array}[t]{r}    -\\    -\end{array}$ & $\begin{array}[t]{r}    -\\     -\end{array}$ &    $-$ &    $-$ &    $-$ &    $-$ & N & N & N & N & N \\
$\begin{array}[t]{c}{\rm MHO}\\1078\end{array}$ & 61.48595   & 51.451418  & $\begin{array}[t]{r}0.00510\\ 0.00327\end{array}$   & $\begin{array}[t]{r}329.1\\161.3\end{array}$ & $\begin{array}[t]{r} 163\\   88\end{array}$ & B & Y & GPS439240249378      & 10 & $\begin{array}[t]{r}13.19\\ 12.26\end{array}$ & $\begin{array}[t]{r}11.59\\11.08\end{array}$ & $\begin{array}[t]{r}10.05\\  9.84\end{array}$ &  8.401 &  7.029 &  4.332 &  1.845 & Y & Y & Y & Y & N \\
$\begin{array}[t]{c}{\rm MHO}\\1079\end{array}$ & 60.98864   & 51.015907  & $\begin{array}[t]{r}0.00241\\ 0.01422\end{array}$   & $\begin{array}[t]{r}348.9\\186.4\end{array}$ & $\begin{array}[t]{r}  80\\   15\end{array}$ & B & Y & GPS439241894456      & 20 & $\begin{array}[t]{r}    -\\     -\end{array}$ & $\begin{array}[t]{r}18.67\\    -\end{array}$ & $\begin{array}[t]{r}16.16\\     -\end{array}$ & 12.678 &  9.685 &  6.912 &  3.818 & Y & N & Y & N & N \\
$\begin{array}[t]{c}{\rm MHO}\\1080\end{array}$ & 61.83245   & 51.360570  & $\begin{array}[t]{r}0.02971\\       -\end{array}$   & $\begin{array}[t]{r} 89.8\\    -\end{array}$ & $\begin{array}[t]{r} 140\\    -\end{array}$ & S & N & J040719.78+512138.0  & 30 & $\begin{array}[t]{r}    -\\     -\end{array}$ & $\begin{array}[t]{r}    -\\    -\end{array}$ & $\begin{array}[t]{r}    -\\     -\end{array}$ & 16.773 & 15.866 & 11.309 &  8.267 & N & N & Y & N & N \\
$\begin{array}[t]{c}{\rm MHO}\\1081\end{array}$ & 61.90227   & 51.265608  & $\begin{array}[t]{r}      -\\       -\end{array}$   & $\begin{array}[t]{r}    -\\    -\end{array}$ & $\begin{array}[t]{r}  20\\    -\end{array}$ & K & Y & unknown              &$-$ & $\begin{array}[t]{r}    -\\     -\end{array}$ & $\begin{array}[t]{r}    -\\    -\end{array}$ & $\begin{array}[t]{r}    -\\     -\end{array}$ &    $-$ &    $-$ &    $-$ &    $-$ & N & N & N & N & N \\
$\begin{array}[t]{c}{\rm MHO}\\1082\end{array}$ & 62.02223   & 51.281981  & $\begin{array}[t]{r}      -\\       -\end{array}$   & $\begin{array}[t]{r}    -\\    -\end{array}$ & $\begin{array}[t]{r}  36\\    -\end{array}$ & K & Y & unknown              &$-$ & $\begin{array}[t]{r}    -\\     -\end{array}$ & $\begin{array}[t]{r}    -\\    -\end{array}$ & $\begin{array}[t]{r}    -\\     -\end{array}$ &    $-$ &    $-$ &    $-$ &    $-$ & N & N & N & N & N \\
$\begin{array}[t]{c}{\rm MHO}\\1083\end{array}$ & 61.45279   & 50.778061  & $\begin{array}[t]{r}0.00813\\       -\end{array}$   & $\begin{array}[t]{r}120.8\\    -\end{array}$ & $\begin{array}[t]{r}  24\\    -\end{array}$ & S & N & GPS439242797745      & 30 & $\begin{array}[t]{r}17.91\\     -\end{array}$ & $\begin{array}[t]{r}16.60\\    -\end{array}$ & $\begin{array}[t]{r}15.74\\     -\end{array}$ & 13.991 & 13.232 & 11.226 &  6.936 & Y & N & Y & N & N \\
$\begin{array}[t]{c}{\rm MHO}\\1084\end{array}$ & 62.27979   & 51.080951  & $\begin{array}[t]{r}0.00764\\       -\end{array}$   & $\begin{array}[t]{r} 51.9\\    -\end{array}$ & $\begin{array}[t]{r}  64\\    -\end{array}$ & S & N & GPS439241917011      & 30 & $\begin{array}[t]{r}19.74\\     -\end{array}$ & $\begin{array}[t]{r}18.28\\    -\end{array}$ & $\begin{array}[t]{r}17.39\\     -\end{array}$ & 15.225 & 12.648 & 10.126 &  6.356 & Y & N & Y & N & N \\
$\begin{array}[t]{c}{\rm MHO}\\1085\end{array}$ & 61.72960   & 50.496437  & $\begin{array}[t]{r}0.00593\\       -\end{array}$   & $\begin{array}[t]{r}  9.3\\    -\end{array}$ & $\begin{array}[t]{r}  16\\    -\end{array}$ & S & N & GPS439244391066      & 30 & $\begin{array}[t]{r}    -\\     -\end{array}$ & $\begin{array}[t]{r}18.45\\    -\end{array}$ & $\begin{array}[t]{r}16.59\\     -\end{array}$ &    $-$ &    $-$ &    $-$ &    $-$ & Y & N & N & N & N \\
$\begin{array}[t]{c}{\rm MHO}\\1086\end{array}$ & 61.71765   & 50.513129  & $\begin{array}[t]{r}0.00162\\ 0.00273\end{array}$   & $\begin{array}[t]{r}359.0\\175.1\end{array}$ & $\begin{array}[t]{r}  48\\  125\end{array}$ & B & N & J040652.23+503047.2  & 90 & $\begin{array}[t]{r}    -\\     -\end{array}$ & $\begin{array}[t]{r}    -\\    -\end{array}$ & $\begin{array}[t]{r}    -\\     -\end{array}$ & 15.569 & 13.013 &  8.887 &  3.659 & N & N & Y & Y & N \\
$\begin{array}[t]{c}{\rm MHO}\\1087\end{array}$ & 61.63636   & 50.507554  & $\begin{array}[t]{r}0.01489\\ 0.00203\end{array}$   & $\begin{array}[t]{r}311.1\\138.8\end{array}$ & $\begin{array}[t]{r}  44\\   19\end{array}$ & B & Y & GPS439244244386      & 20 & $\begin{array}[t]{r}    -\\     -\end{array}$ & $\begin{array}[t]{r}    -\\    -\end{array}$ & $\begin{array}[t]{r}16.59\\     -\end{array}$ & 12.653 & 10.291 &  6.706 &  3.093 & Y & N & Y & Y & N \\
$\begin{array}[t]{c}{\rm MHO}\\1088\end{array}$ & 61.87309   & 50.530046  & $\begin{array}[t]{r}0.00208\\ 0.00108\end{array}$   & $\begin{array}[t]{r} 46.8\\231.3\end{array}$ & $\begin{array}[t]{r}  23\\   69\end{array}$ & B & N & J040729.54+503148.1  & 90 & $\begin{array}[t]{r}    -\\     -\end{array}$ & $\begin{array}[t]{r}    -\\    -\end{array}$ & $\begin{array}[t]{r}    -\\     -\end{array}$ & 16.700 & 14.942 & 12.078 &  9.008 & N & N & Y & N & N \\
$\begin{array}[t]{c}{\rm MHO}\\1089\end{array}$ & 62.00873   & 50.538308  & $\begin{array}[t]{r}      -\\       -\end{array}$   & $\begin{array}[t]{r}    -\\    -\end{array}$ & $\begin{array}[t]{r}  26\\    -\end{array}$ & K & Y & unknown              &$-$ & $\begin{array}[t]{r}    -\\     -\end{array}$ & $\begin{array}[t]{r}    -\\    -\end{array}$ & $\begin{array}[t]{r}    -\\     -\end{array}$ &    $-$ &    $-$ &    $-$ &    $-$ & N & N & N & N & N \\
$\begin{array}[t]{c}{\rm MHO}\\1090\end{array}$ & 62.02896   & 50.523539  & $\begin{array}[t]{r}      -\\       -\end{array}$   & $\begin{array}[t]{r}    -\\    -\end{array}$ & $\begin{array}[t]{r}  34\\    -\end{array}$ & K & Y & unknown              &$-$ & $\begin{array}[t]{r}    -\\     -\end{array}$ & $\begin{array}[t]{r}    -\\    -\end{array}$ & $\begin{array}[t]{r}    -\\     -\end{array}$ &    $-$ &    $-$ &    $-$ &    $-$ & N & N & N & N & N \\
$\begin{array}[t]{c}{\rm MHO}\\1091\end{array}$ & 62.04327   & 50.522287  & $\begin{array}[t]{r}0.00199\\ 0.00393\end{array}$   & $\begin{array}[t]{r}298.4\\125.0\end{array}$ & $\begin{array}[t]{r}  50\\   18\end{array}$ & B & Y & GPS439244580755      & 60 & $\begin{array}[t]{r}16.53\\ 16.94\end{array}$ & $\begin{array}[t]{r}14.21\\14.85\end{array}$ & $\begin{array}[t]{r}12.29\\ 12.75\end{array}$ & 10.749 &  9.534 &  6.540 &  3.538 & Y & Y & Y & Y & N \\
$\begin{array}[t]{c}{\rm MHO}\\1092\end{array}$ & 61.36135   & 49.652435  & $\begin{array}[t]{r}0.00523\\ 0.00810\end{array}$   & $\begin{array}[t]{r} 89.1\\262.9\end{array}$ & $\begin{array}[t]{r} 152\\  476\end{array}$ & B & Y & GPS439249569784      & 20 & $\begin{array}[t]{r}    -\\     -\end{array}$ & $\begin{array}[t]{r}18.52\\    -\end{array}$ & $\begin{array}[t]{r}16.86\\     -\end{array}$ &    $-$ &    $-$ &    $-$ &    $-$ & Y & N & N & N & N \\
$\begin{array}[t]{c}{\rm MHO}\\1093\end{array}$ & 61.36233   & 49.650729  & $\begin{array}[t]{r}0.00664\\ 0.00463\end{array}$   & $\begin{array}[t]{r}  7.3\\185.8\end{array}$ & $\begin{array}[t]{r}  41\\   27\end{array}$ & B & Y & GPS439249565961      & 20 & $\begin{array}[t]{r}19.19\\     -\end{array}$ & $\begin{array}[t]{r}17.15\\    -\end{array}$ & $\begin{array}[t]{r}14.81\\     -\end{array}$ & 11.931 & 10.047 &  7.495 &  4.368 & Y & N & Y & N & N \\
$\begin{array}[t]{c}{\rm MHO}\\1094\end{array}$ & 61.36369   & 49.651530  & $\begin{array}[t]{r}0.00840\\ 0.00671\end{array}$   & $\begin{array}[t]{r} 26.5\\212.2\end{array}$ & $\begin{array}[t]{r}  24\\   72\end{array}$ & B & Y & GPS439249565937      & 20 & $\begin{array}[t]{r}18.40\\ 18.17\end{array}$ & $\begin{array}[t]{r}16.09\\15.71\end{array}$ & $\begin{array}[t]{r}14.50\\ 14.30\end{array}$ & 11.931 & 10.047 &  7.495 &  4.368 & Y & Y & Y & N & N \\
$\begin{array}[t]{c}{\rm MHO}\\1095\end{array}$ & 62.17080   & 49.795180  & $\begin{array}[t]{r}      -\\       -\end{array}$   & $\begin{array}[t]{r}    -\\    -\end{array}$ & $\begin{array}[t]{r}  39\\    -\end{array}$ & K & N & unknown              &$-$ & $\begin{array}[t]{r}    -\\     -\end{array}$ & $\begin{array}[t]{r}    -\\    -\end{array}$ & $\begin{array}[t]{r}    -\\     -\end{array}$ &    $-$ &    $-$ &    $-$ &    $-$ & N & N & N & N & N \\
$\begin{array}[t]{c}{\rm MHO}\\1096\end{array}$ & 63.33788   & 50.054994  & $\begin{array}[t]{r}      -\\       -\end{array}$   & $\begin{array}[t]{r}    -\\    -\end{array}$ & $\begin{array}[t]{r}  50\\    -\end{array}$ & K & Y & unknown              &$-$ & $\begin{array}[t]{r}    -\\     -\end{array}$ & $\begin{array}[t]{r}    -\\    -\end{array}$ & $\begin{array}[t]{r}    -\\     -\end{array}$ &    $-$ &    $-$ &    $-$ &    $-$ & N & N & N & N & N \\
$\begin{array}[t]{c}{\rm MHO}\\1097\end{array}$ & 65.56849   & 50.571718  & $\begin{array}[t]{r}0.00257\\ 0.00288\end{array}$   & $\begin{array}[t]{r}333.7\\157.6\end{array}$ & $\begin{array}[t]{r}  22\\  107\end{array}$ & B & N & GPS439243601173      & 30 & $\begin{array}[t]{r}19.84\\     -\end{array}$ & $\begin{array}[t]{r}    -\\    -\end{array}$ & $\begin{array}[t]{r}17.24\\     -\end{array}$ &    $-$ &    $-$ &    $-$ &    $-$ & Y & N & N & N & N \\
$\begin{array}[t]{c}{\rm MHO}\\1098\end{array}$ & 65.57517   & 50.568972  & $\begin{array}[t]{r}      -\\       -\end{array}$   & $\begin{array}[t]{r}    -\\    -\end{array}$ & $\begin{array}[t]{r}  27\\    -\end{array}$ & K & N & unknown              &$-$ & $\begin{array}[t]{r}    -\\     -\end{array}$ & $\begin{array}[t]{r}    -\\    -\end{array}$ & $\begin{array}[t]{r}    -\\     -\end{array}$ &    $-$ &    $-$ &    $-$ &    $-$ & N & N & N & N & N \\
$\begin{array}[t]{c}{\rm MHO}\\1099\end{array}$ & 65.46749   & 50.606260  & $\begin{array}[t]{r}0.00325\\ 0.00425\end{array}$   & $\begin{array}[t]{r} 23.7\\203.9\end{array}$ & $\begin{array}[t]{r}  19\\   48\end{array}$ & B & N & GPS439243608737      & 50 & $\begin{array}[t]{r}19.51\\     -\end{array}$ & $\begin{array}[t]{r}17.69\\    -\end{array}$ & $\begin{array}[t]{r}16.73\\     -\end{array}$ & 13.444 & 10.707 &  7.959 &  4.485 & Y & N & Y & Y & Y \\
$\begin{array}[t]{c}{\rm MHO}\\2978\end{array}$ & 47.06005   & 56.776384  & $\begin{array}[t]{r}      -\\       -\end{array}$   & $\begin{array}[t]{r}    -\\    -\end{array}$ & $\begin{array}[t]{r}  24\\    -\end{array}$ & K & N & unknown              &$-$ & $\begin{array}[t]{r}    -\\     -\end{array}$ & $\begin{array}[t]{r}    -\\    -\end{array}$ & $\begin{array}[t]{r}    -\\     -\end{array}$ &    $-$ &    $-$ &    $-$ &    $-$ & N & N & N & N & N \\
$\begin{array}[t]{c}{\rm MHO}\\2979\end{array}$ & 47.02932   & 56.766349  & $\begin{array}[t]{r}0.00979\\ 0.01112\end{array}$   & $\begin{array}[t]{r}  9.1\\198.4\end{array}$ & $\begin{array}[t]{r}  64\\   89\end{array}$ & B & N & GPS439222446979      & 90 & $\begin{array}[t]{r}19.16\\ 18.89\end{array}$ & $\begin{array}[t]{r}16.95\\16.19\end{array}$ & $\begin{array}[t]{r}14.77\\ 14.47\end{array}$ & 12.942 & 10.997 &  8.782 &  5.788 & Y & Y & Y & N & N \\
$\begin{array}[t]{c}{\rm MHO}\\2980\end{array}$ & 47.07636   & 56.393501  & $\begin{array}[t]{r}0.01199\\       -\end{array}$   & $\begin{array}[t]{r}335.4\\    -\end{array}$ & $\begin{array}[t]{r}  49\\    -\end{array}$ & S & N & GPS439223566844      & 90 & $\begin{array}[t]{r}16.75\\ 14.98\end{array}$ & $\begin{array}[t]{r}13.88\\12.95\end{array}$ & $\begin{array}[t]{r}11.05\\ 10.75\end{array}$ &  8.270 &  6.473 &  3.314 &  0.739 & Y & Y & Y & N & N \\
$\begin{array}[t]{c}{\rm MHO}\\2981\end{array}$ & 51.94579   & 58.977700  & $\begin{array}[t]{r}0.00995\\ 0.00693\end{array}$   & $\begin{array}[t]{r}112.3\\308.2\end{array}$ & $\begin{array}[t]{r}  46\\  187\end{array}$ & B & N & GPS438826850299      & 40 & $\begin{array}[t]{r}17.64\\     -\end{array}$ & $\begin{array}[t]{r}16.69\\    -\end{array}$ & $\begin{array}[t]{r}16.12\\     -\end{array}$ & 16.106 & 16.956 & 10.599 &  8.744 & Y & N & Y & N & N \\
$\begin{array}[t]{c}{\rm MHO}\\2982\end{array}$ & 51.86793   & 58.903200  & $\begin{array}[t]{r}0.01807\\       -\end{array}$   & $\begin{array}[t]{r} 79.1\\    -\end{array}$ & $\begin{array}[t]{r}1074\\    -\end{array}$ & S & Y & GPS438826615365      & 90 & $\begin{array}[t]{r}20.20\\ 16.67\end{array}$ & $\begin{array}[t]{r}18.14\\15.27\end{array}$ & $\begin{array}[t]{r}15.66\\ 14.34\end{array}$ & 11.870 &  9.856 &  5.882 &  0.574 & Y & Y & Y & N & Y \\
$\begin{array}[t]{c}{\rm MHO}\\2983\end{array}$ & 50.73902   & 55.051985  & $\begin{array}[t]{r}0.00542\\       -\end{array}$   & $\begin{array}[t]{r}155.4\\    -\end{array}$ & $\begin{array}[t]{r}  32\\    -\end{array}$ & S & N & GPS439226436484      & 60 & $\begin{array}[t]{r}    -\\     -\end{array}$ & $\begin{array}[t]{r}    -\\    -\end{array}$ & $\begin{array}[t]{r}17.46\\     -\end{array}$ & 15.266 & 13.278 & 11.248 &  5.136 & Y & N & Y & Y & N \\
$\begin{array}[t]{c}{\rm MHO}\\2984\end{array}$ & 50.65753   & 55.060378  & $\begin{array}[t]{r}0.00186\\       -\end{array}$   & $\begin{array}[t]{r} 15.6\\    -\end{array}$ & $\begin{array}[t]{r}  26\\    -\end{array}$ & S & Y & GPS439226432804      & 30 & $\begin{array}[t]{r}17.79\\ 16.30\end{array}$ & $\begin{array}[t]{r}15.22\\14.73\end{array}$ & $\begin{array}[t]{r}13.43\\ 13.16\end{array}$ & 10.261 &  8.720 &  6.349 &  2.724 & Y & Y & Y & Y & N \\
$\begin{array}[t]{c}{\rm MHO}\\2985\end{array}$ & 50.65956   & 55.061489  & $\begin{array}[t]{r}0.03486\\ 0.01589\end{array}$   & $\begin{array}[t]{r}115.1\\295.8\end{array}$ & $\begin{array}[t]{r} 105\\  118\end{array}$ & B & Y & GPS439226432818      & 20 & $\begin{array}[t]{r}17.14\\ 16.54\end{array}$ & $\begin{array}[t]{r}14.44\\14.32\end{array}$ & $\begin{array}[t]{r}12.28\\ 12.39\end{array}$ & 11.261 &  9.720 &  7.349 &  3.724 & Y & Y & Y & Y & N \\
$\begin{array}[t]{c}{\rm MHO}\\2986\end{array}$ & 51.11426   & 55.203083  & $\begin{array}[t]{r}0.00325\\ 0.00371\end{array}$   & $\begin{array}[t]{r}158.7\\338.1\end{array}$ & $\begin{array}[t]{r} 475\\   78\end{array}$ & B & Y & GPS439225653298      & 10 & $\begin{array}[t]{r}    -\\     -\end{array}$ & $\begin{array}[t]{r}17.39\\    -\end{array}$ & $\begin{array}[t]{r}14.85\\     -\end{array}$ &    $-$ &    $-$ &    $-$ &    $-$ & Y & N & N & N & N \\
$\begin{array}[t]{c}{\rm MHO}\\2987\end{array}$ & 51.11123   & 55.210352  & $\begin{array}[t]{r}      -\\       -\end{array}$   & $\begin{array}[t]{r}    -\\    -\end{array}$ & $\begin{array}[t]{r}  22\\    -\end{array}$ & K & Y & unknown              &$-$ & $\begin{array}[t]{r}    -\\     -\end{array}$ & $\begin{array}[t]{r}    -\\    -\end{array}$ & $\begin{array}[t]{r}    -\\     -\end{array}$ &    $-$ &    $-$ &    $-$ &    $-$ & N & N & N & N & N \\
$\begin{array}[t]{c}{\rm MHO}\\2988\end{array}$ & 51.09660   & 55.210308  & $\begin{array}[t]{r}      -\\       -\end{array}$   & $\begin{array}[t]{r}    -\\    -\end{array}$ & $\begin{array}[t]{r}  38\\    -\end{array}$ & K & Y & unknown              &$-$ & $\begin{array}[t]{r}    -\\     -\end{array}$ & $\begin{array}[t]{r}    -\\    -\end{array}$ & $\begin{array}[t]{r}    -\\     -\end{array}$ &    $-$ &    $-$ &    $-$ &    $-$ & N & N & N & N & N \\
$\begin{array}[t]{c}{\rm MHO}\\2989\end{array}$ & 50.98008   & 54.849785  & $\begin{array}[t]{r}0.00501\\       -\end{array}$   & $\begin{array}[t]{r}299.9\\    -\end{array}$ & $\begin{array}[t]{r} 103\\    -\end{array}$ & S & N & GPS439227089728      & 20 & $\begin{array}[t]{r}    -\\     -\end{array}$ & $\begin{array}[t]{r}    -\\    -\end{array}$ & $\begin{array}[t]{r}17.76\\     -\end{array}$ &    $-$ &    $-$ &    $-$ &    $-$ & Y & N & N & N & Y \\
$\begin{array}[t]{c}{\rm MHO}\\2990\end{array}$ & 51.45892   & 55.195615  & $\begin{array}[t]{r}0.00742\\ 0.01030\end{array}$   & $\begin{array}[t]{r}286.8\\115.1\end{array}$ & $\begin{array}[t]{r}  71\\   66\end{array}$ & B & N & GPS439225484218      & 90 & $\begin{array}[t]{r}17.96\\ 15.90\end{array}$ & $\begin{array}[t]{r}15.43\\15.42\end{array}$ & $\begin{array}[t]{r}13.69\\ 13.71\end{array}$ & 12.373 & 10.866 &  8.257 &  5.192 & Y & Y & Y & Y & N \\
$\begin{array}[t]{c}{\rm MHO}\\2991\end{array}$ & 51.23580   & 54.938369  & $\begin{array}[t]{r}0.00832\\ 0.00809\end{array}$   & $\begin{array}[t]{r}308.8\\135.8\end{array}$ & $\begin{array}[t]{r} 136\\  412\end{array}$ & B & Y & GPS439227083972      & 90 & $\begin{array}[t]{r}19.09\\ 18.67\end{array}$ & $\begin{array}[t]{r}16.53\\16.02\end{array}$ & $\begin{array}[t]{r}14.22\\ 14.01\end{array}$ & 10.856 &  9.243 &  5.171 &  2.878 & Y & Y & Y & N & N \\
$\begin{array}[t]{c}{\rm MHO}\\2992\end{array}$ & 51.28495   & 54.956925  & $\begin{array}[t]{r}      -\\       -\end{array}$   & $\begin{array}[t]{r}    -\\    -\end{array}$ & $\begin{array}[t]{r}  46\\    -\end{array}$ & K & Y & unknown              &$-$ & $\begin{array}[t]{r}    -\\     -\end{array}$ & $\begin{array}[t]{r}    -\\    -\end{array}$ & $\begin{array}[t]{r}    -\\     -\end{array}$ &    $-$ &    $-$ &    $-$ &    $-$ & N & N & N & N & N \\
$\begin{array}[t]{c}{\rm MHO}\\2993\end{array}$ & 51.25459   & 54.979435  & $\begin{array}[t]{r}0.00109\\ 0.00081\end{array}$   & $\begin{array}[t]{r} 39.5\\219.1\end{array}$ & $\begin{array}[t]{r}  66\\   30\end{array}$ & B & N & J032501.11+545845.8  & 90 & $\begin{array}[t]{r}    -\\     -\end{array}$ & $\begin{array}[t]{r}    -\\    -\end{array}$ & $\begin{array}[t]{r}    -\\     -\end{array}$ & 14.289 & 11.809 &  7.805 &  3.861 & N & N & Y & N & N \\
$\begin{array}[t]{c}{\rm MHO}\\2994\end{array}$ & 52.14302   & 55.180613  & $\begin{array}[t]{r}0.00139\\ 0.00139\end{array}$   & $\begin{array}[t]{r}321.8\\148.1\end{array}$ & $\begin{array}[t]{r}  18\\   18\end{array}$ & B & N & GPS439225550906      & 90 & $\begin{array}[t]{r}17.60\\ 16.91\end{array}$ & $\begin{array}[t]{r}15.60\\15.18\end{array}$ & $\begin{array}[t]{r}14.05\\ 13.79\end{array}$ & 12.542 & 10.928 &  8.259 &  4.939 & Y & Y & Y & N & N \\
$\begin{array}[t]{c}{\rm MHO}\\2995\end{array}$ & 51.61656   & 54.682123  & $\begin{array}[t]{r}0.01190\\ 0.00763\end{array}$   & $\begin{array}[t]{r}321.3\\132.5\end{array}$ & $\begin{array}[t]{r}  32\\  187\end{array}$ & B & Y & GPS439228061107      & 30 & $\begin{array}[t]{r}17.10\\ 17.60\end{array}$ & $\begin{array}[t]{r}14.59\\15.55\end{array}$ & $\begin{array}[t]{r}12.48\\ 14.01\end{array}$ & 10.179 &  8.415 &  5.384 &  2.629 & Y & Y & Y & Y & Y \\
$\begin{array}[t]{c}{\rm MHO}\\2996\end{array}$ & 52.98990   & 55.645423  & $\begin{array}[t]{r}0.00195\\       -\end{array}$   & $\begin{array}[t]{r} 90.9\\    -\end{array}$ & $\begin{array}[t]{r} 384\\    -\end{array}$ & S & Y & GPS439224518813      & 20 & $\begin{array}[t]{r}14.69\\ 14.70\end{array}$ & $\begin{array}[t]{r}13.44\\13.37\end{array}$ & $\begin{array}[t]{r}12.70\\ 12.65\end{array}$ & 11.282 & 10.510 &  7.244 &  5.039 & Y & Y & Y & N & N \\
$\begin{array}[t]{c}{\rm MHO}\\2997\end{array}$ & 52.99320   & 55.645900  & $\begin{array}[t]{r}0.00573\\ 0.00736\end{array}$   & $\begin{array}[t]{r}326.1\\150.9\end{array}$ & $\begin{array}[t]{r}  14\\   19\end{array}$ & B & Y & 0331584+553845       & 50 & $\begin{array}[t]{r}    -\\     -\end{array}$ & $\begin{array}[t]{r}    -\\    -\end{array}$ & $\begin{array}[t]{r}    -\\     -\end{array}$ &    $-$ &    $-$ &    $-$ &    $-$ & N & N & N & Y & N \\
$\begin{array}[t]{c}{\rm MHO}\\2998\end{array}$ & 52.30408   & 54.716561  & $\begin{array}[t]{r}      -\\       -\end{array}$   & $\begin{array}[t]{r}    -\\    -\end{array}$ & $\begin{array}[t]{r}  15\\    -\end{array}$ & K & N & unknown              &$-$ & $\begin{array}[t]{r}    -\\     -\end{array}$ & $\begin{array}[t]{r}    -\\    -\end{array}$ & $\begin{array}[t]{r}    -\\     -\end{array}$ &    $-$ &    $-$ &    $-$ &    $-$ & N & N & N & N & N \\
$\begin{array}[t]{c}{\rm MHO}\\2999\end{array}$ & 52.88295   & 55.100457  & $\begin{array}[t]{r}0.00168\\       -\end{array}$   & $\begin{array}[t]{r} 69.9\\    -\end{array}$ & $\begin{array}[t]{r}  41\\    -\end{array}$ & S & N & J033131.90+550601.6  & 70 & $\begin{array}[t]{r}    -\\     -\end{array}$ & $\begin{array}[t]{r}    -\\    -\end{array}$ & $\begin{array}[t]{r}    -\\     -\end{array}$ & 16.769 & 15.099 & 12.451 &  8.930 & N & N & Y & N & N \\
$\begin{array}[t]{c}{\rm MHO}\\3600\end{array}$ & 64.81005   & 48.418351  & $\begin{array}[t]{r}0.00184\\ 0.00222\end{array}$   & $\begin{array}[t]{r}101.3\\279.1\end{array}$ & $\begin{array}[t]{r}   9\\   12\end{array}$ & B & N & GPS438834402010      & 90 & $\begin{array}[t]{r}    -\\     -\end{array}$ & $\begin{array}[t]{r}17.85\\    -\end{array}$ & $\begin{array}[t]{r}15.81\\     -\end{array}$ & 14.185 & 11.977 &  9.048 &  5.982 & Y & N & Y & N & N \\
$\begin{array}[t]{c}{\rm MHO}\\3601\end{array}$ & 64.80394   & 48.433302  & $\begin{array}[t]{r}      -\\       -\end{array}$   & $\begin{array}[t]{r}    -\\    -\end{array}$ & $\begin{array}[t]{r}  52\\    -\end{array}$ & K & N & unknown              &$-$ & $\begin{array}[t]{r}    -\\     -\end{array}$ & $\begin{array}[t]{r}    -\\    -\end{array}$ & $\begin{array}[t]{r}    -\\     -\end{array}$ &    $-$ &    $-$ &    $-$ &    $-$ & N & N & N & N & N \\
$\begin{array}[t]{c}{\rm MHO}\\3602\end{array}$ & 64.83847   & 48.420403  & $\begin{array}[t]{r}      -\\       -\end{array}$   & $\begin{array}[t]{r}    -\\    -\end{array}$ & $\begin{array}[t]{r}  19\\    -\end{array}$ & K & N & unknown              &$-$ & $\begin{array}[t]{r}    -\\     -\end{array}$ & $\begin{array}[t]{r}    -\\    -\end{array}$ & $\begin{array}[t]{r}    -\\     -\end{array}$ &    $-$ &    $-$ &    $-$ &    $-$ & N & N & N & N & N \\
$\begin{array}[t]{c}{\rm MHO}\\3700\end{array}$ & 53.29372   & 55.167976  & $\begin{array}[t]{r}0.00181\\ 0.00421\end{array}$   & $\begin{array}[t]{r} 53.7\\241.4\end{array}$ & $\begin{array}[t]{r}  19\\   47\end{array}$ & B & Y & GPS439226302222      & 30 & $\begin{array}[t]{r}17.34\\     -\end{array}$ & $\begin{array}[t]{r}16.32\\    -\end{array}$ & $\begin{array}[t]{r}15.77\\     -\end{array}$ & 15.094 & 14.385 & 10.848 &  6.338 & Y & N & Y & N & N \\
$\begin{array}[t]{c}{\rm MHO}\\3701\end{array}$ & 53.28411   & 55.171591  & $\begin{array}[t]{r}0.00214\\ 0.00198\end{array}$   & $\begin{array}[t]{r} 57.6\\239.4\end{array}$ & $\begin{array}[t]{r}  69\\   28\end{array}$ & B & Y & GPS439225471182      & 40 & $\begin{array}[t]{r}14.00\\ 13.98\end{array}$ & $\begin{array}[t]{r}12.90\\12.64\end{array}$ & $\begin{array}[t]{r}11.96\\ 11.77\end{array}$ & 10.326 &  8.940 &  5.771 &  3.031 & Y & Y & Y & Y & N \\
$\begin{array}[t]{c}{\rm MHO}\\3702\end{array}$ & 53.28489   & 55.175693  & $\begin{array}[t]{r}0.00491\\ 0.00119\end{array}$   & $\begin{array}[t]{r}328.1\\150.0\end{array}$ & $\begin{array}[t]{r}  43\\  158\end{array}$ & B & Y & GPS439225471259      & 20 & $\begin{array}[t]{r}18.05\\     -\end{array}$ & $\begin{array}[t]{r}17.10\\    -\end{array}$ & $\begin{array}[t]{r}16.53\\     -\end{array}$ &    $-$ &    $-$ &    $-$ &    $-$ & Y & N & N & N & N \\
$\begin{array}[t]{c}{\rm MHO}\\3703\end{array}$ & 53.12653   & 54.845795  & $\begin{array}[t]{r}0.00768\\ 0.00865\end{array}$   & $\begin{array}[t]{r}124.8\\312.0\end{array}$ & $\begin{array}[t]{r} 128\\   74\end{array}$ & B & N & GPS439227227879      & 70 & $\begin{array}[t]{r}14.35\\ 14.63\end{array}$ & $\begin{array}[t]{r}13.31\\13.40\end{array}$ & $\begin{array}[t]{r}12.06\\ 12.22\end{array}$ & 10.371 &  9.136 &  6.441 &  4.273 & Y & Y & Y & Y & N \\
$\begin{array}[t]{c}{\rm MHO}\\3704\end{array}$ & 53.03394   & 54.901486  & $\begin{array}[t]{r}0.00110\\ 0.00169\end{array}$   & $\begin{array}[t]{r}316.7\\133.7\end{array}$ & $\begin{array}[t]{r}  47\\   58\end{array}$ & B & N & J033208.14+545405.3  & 90 & $\begin{array}[t]{r}    -\\     -\end{array}$ & $\begin{array}[t]{r}    -\\    -\end{array}$ & $\begin{array}[t]{r}    -\\     -\end{array}$ & 16.278 & 14.704 & 12.372 &  9.117 & N & N & Y & N & N \\
$\begin{array}[t]{c}{\rm MHO}\\3705\end{array}$ & 52.89732   & 54.845318  & $\begin{array}[t]{r}0.00213\\ 0.00234\end{array}$   & $\begin{array}[t]{r} 89.6\\264.5\end{array}$ & $\begin{array}[t]{r}  18\\   10\end{array}$ & B & N & noname               & 10 & $\begin{array}[t]{r}    -\\     -\end{array}$ & $\begin{array}[t]{r}    -\\    -\end{array}$ & $\begin{array}[t]{r}    -\\     -\end{array}$ &    $-$ &    $-$ &    $-$ &    $-$ & N & N & N & N & N \\
$\begin{array}[t]{c}{\rm MHO}\\3706\end{array}$ & 55.36815   & 56.228536  & $\begin{array}[t]{r}0.00110\\ 0.00566\end{array}$   & $\begin{array}[t]{r}263.0\\ 68.3\end{array}$ & $\begin{array}[t]{r}  54\\   28\end{array}$ & B & N & J034128.35+561342.7  & 50 & $\begin{array}[t]{r}    -\\     -\end{array}$ & $\begin{array}[t]{r}    -\\    -\end{array}$ & $\begin{array}[t]{r}    -\\     -\end{array}$ & 16.757 & 15.049 & 12.242 &  9.153 & N & N & Y & N & N \\
$\begin{array}[t]{c}{\rm MHO}\\3707\end{array}$ & 56.29465   & 54.526851  & $\begin{array}[t]{r}0.00312\\ 0.00304\end{array}$   & $\begin{array}[t]{r}115.2\\303.5\end{array}$ & $\begin{array}[t]{r}  80\\   65\end{array}$ & B & Y & noname               & 90 & $\begin{array}[t]{r}    -\\     -\end{array}$ & $\begin{array}[t]{r}    -\\    -\end{array}$ & $\begin{array}[t]{r}    -\\     -\end{array}$ &    $-$ &    $-$ &    $-$ &    $-$ & N & N & N & N & N \\
$\begin{array}[t]{c}{\rm MHO}\\3708\end{array}$ & 58.77886   & 53.798940  & $\begin{array}[t]{r}0.00167\\ 0.00248\end{array}$   & $\begin{array}[t]{r} 95.7\\274.7\end{array}$ & $\begin{array}[t]{r}  15\\    8\end{array}$ & B & N & J035506.92+534756.1  & 50 & $\begin{array}[t]{r}    -\\     -\end{array}$ & $\begin{array}[t]{r}18.22\\    -\end{array}$ & $\begin{array}[t]{r}16.14\\     -\end{array}$ & 14.613 & 12.354 & 10.857 &  6.488 & Y & N & Y & N & N \\
$\begin{array}[t]{c}{\rm MHO}\\3709\end{array}$ & 58.87122   & 53.800635  & $\begin{array}[t]{r}0.00664\\ 0.01009\end{array}$   & $\begin{array}[t]{r} 92.2\\269.2\end{array}$ & $\begin{array}[t]{r}  72\\   55\end{array}$ & B & N & noname               & 90 & $\begin{array}[t]{r}    -\\     -\end{array}$ & $\begin{array}[t]{r}    -\\    -\end{array}$ & $\begin{array}[t]{r}    -\\     -\end{array}$ & 15.665 & 13.532 & 10.359 &  5.410 & N & N & Y & N & N \\
$\begin{array}[t]{c}{\rm MHO}\\3710\end{array}$ & 58.88664   & 53.747850  & $\begin{array}[t]{r}0.00158\\ 0.00193\end{array}$   & $\begin{array}[t]{r} 91.9\\271.1\end{array}$ & $\begin{array}[t]{r}  73\\   46\end{array}$ & B & N & GPS439231466690      & 90 & $\begin{array}[t]{r}    -\\     -\end{array}$ & $\begin{array}[t]{r}18.72\\    -\end{array}$ & $\begin{array}[t]{r}17.45\\     -\end{array}$ & 16.434 & 14.632 & 11.334 &  7.230 & Y & N & Y & N & N \\
$\begin{array}[t]{c}{\rm MHO}\\3711\end{array}$ & 58.88604   & 53.741337  & $\begin{array}[t]{r}0.00078\\ 0.00073\end{array}$   & $\begin{array}[t]{r} 59.4\\235.3\end{array}$ & $\begin{array}[t]{r}  12\\   16\end{array}$ & B & N & J035532.60+534428.9  & 90 & $\begin{array}[t]{r}    -\\     -\end{array}$ & $\begin{array}[t]{r}    -\\    -\end{array}$ & $\begin{array}[t]{r}    -\\     -\end{array}$ & 16.064 & 14.687 & 11.792 &  8.314 & N & N & Y & N & N \\
$\begin{array}[t]{c}{\rm MHO}\\3712\end{array}$ & 58.88308   & 53.776865  & $\begin{array}[t]{r}      -\\       -\end{array}$   & $\begin{array}[t]{r}    -\\    -\end{array}$ & $\begin{array}[t]{r} 163\\    -\end{array}$ & K & N & unknown              &$-$ & $\begin{array}[t]{r}    -\\     -\end{array}$ & $\begin{array}[t]{r}    -\\    -\end{array}$ & $\begin{array}[t]{r}    -\\     -\end{array}$ &    $-$ &    $-$ &    $-$ &    $-$ & N & N & N & N & N \\
$\begin{array}[t]{c}{\rm MHO}\\3713\end{array}$ & 58.88825   & 53.767608  & $\begin{array}[t]{r}      -\\       -\end{array}$   & $\begin{array}[t]{r}    -\\    -\end{array}$ & $\begin{array}[t]{r} 221\\    -\end{array}$ & K & N & unknown              &$-$ & $\begin{array}[t]{r}    -\\     -\end{array}$ & $\begin{array}[t]{r}    -\\    -\end{array}$ & $\begin{array}[t]{r}    -\\     -\end{array}$ &    $-$ &    $-$ &    $-$ &    $-$ & N & N & N & N & N \\
$\begin{array}[t]{c}{\rm MHO}\\3714\end{array}$ & 58.87778   & 53.763318  & $\begin{array}[t]{r}0.00184\\       -\end{array}$   & $\begin{array}[t]{r} 41.2\\    -\end{array}$ & $\begin{array}[t]{r} 123\\    -\end{array}$ & S & N & GPS439231463431      & 30 & $\begin{array}[t]{r}19.34\\     -\end{array}$ & $\begin{array}[t]{r}18.00\\    -\end{array}$ & $\begin{array}[t]{r}16.80\\     -\end{array}$ &    $-$ &    $-$ &    $-$ &    $-$ & Y & N & N & N & N \\
$\begin{array}[t]{c}{\rm MHO}\\3715\end{array}$ & 58.85462   & 53.751984  & $\begin{array}[t]{r}0.00387\\ 0.00289\end{array}$   & $\begin{array}[t]{r}102.0\\281.8\end{array}$ & $\begin{array}[t]{r}  63\\   24\end{array}$ & B & N & noname               & 30 & $\begin{array}[t]{r}    -\\     -\end{array}$ & $\begin{array}[t]{r}    -\\    -\end{array}$ & $\begin{array}[t]{r}    -\\     -\end{array}$ &    $-$ &    $-$ &    $-$ &    $-$ & N & N & N & N & N \\
$\begin{array}[t]{c}{\rm MHO}\\3716\end{array}$ & 59.27117   & 53.781567  & $\begin{array}[t]{r}0.00095\\ 0.00083\end{array}$   & $\begin{array}[t]{r} 43.7\\229.4\end{array}$ & $\begin{array}[t]{r}   8\\   11\end{array}$ & B & N & J035705.13+534654.0  & 30 & $\begin{array}[t]{r}    -\\     -\end{array}$ & $\begin{array}[t]{r}    -\\    -\end{array}$ & $\begin{array}[t]{r}    -\\     -\end{array}$ & 17.474 & 15.323 & 11.830 &  9.092 & N & N & Y & N & N \\
$\begin{array}[t]{c}{\rm MHO}\\3717\end{array}$ & 59.15515   & 53.800516  & $\begin{array}[t]{r}0.00042\\ 0.00070\end{array}$   & $\begin{array}[t]{r} 12.1\\198.8\end{array}$ & $\begin{array}[t]{r}  12\\   17\end{array}$ & B & N & GPS439231567760      & 90 & $\begin{array}[t]{r}    -\\     -\end{array}$ & $\begin{array}[t]{r}19.26\\    -\end{array}$ & $\begin{array}[t]{r}16.57\\     -\end{array}$ & 14.229 & 11.913 &  8.924 &  3.898 & Y & N & Y & Y & N \\
$\begin{array}[t]{c}{\rm MHO}\\3718\end{array}$ & 59.05078   & 53.797462  & $\begin{array}[t]{r}0.00631\\ 0.00229\end{array}$   & $\begin{array}[t]{r} 52.1\\239.5\end{array}$ & $\begin{array}[t]{r}  72\\  109\end{array}$ & B & N & GPS439231563437      & 30 & $\begin{array}[t]{r}18.35\\     -\end{array}$ & $\begin{array}[t]{r}16.89\\    -\end{array}$ & $\begin{array}[t]{r}15.44\\     -\end{array}$ & 13.951 & 12.134 &  8.746 &  5.188 & Y & N & Y & N & N \\
$\begin{array}[t]{c}{\rm MHO}\\3719\end{array}$ & 59.01931   & 53.819373  & $\begin{array}[t]{r}0.00162\\       -\end{array}$   & $\begin{array}[t]{r}217.5\\    -\end{array}$ & $\begin{array}[t]{r}  55\\    -\end{array}$ & S & N & GPS439231563011      & 30 & $\begin{array}[t]{r}16.13\\ 16.16\end{array}$ & $\begin{array}[t]{r}15.38\\15.43\end{array}$ & $\begin{array}[t]{r}14.95\\ 14.95\end{array}$ & 14.574 & 14.057 & 11.438 &  7.815 & Y & Y & Y & N & N \\
$\begin{array}[t]{c}{\rm MHO}\\3720\end{array}$ & 59.06760   & 53.844351  & $\begin{array}[t]{r}0.00190\\ 0.00198\end{array}$   & $\begin{array}[t]{r} 19.9\\194.4\end{array}$ & $\begin{array}[t]{r} 211\\  189\end{array}$ & B & N & noname               & 50 & $\begin{array}[t]{r}    -\\     -\end{array}$ & $\begin{array}[t]{r}    -\\    -\end{array}$ & $\begin{array}[t]{r}    -\\     -\end{array}$ &    $-$ &    $-$ &    $-$ &    $-$ & N & N & N & N & N \\
$\begin{array}[t]{c}{\rm MHO}\\3721\end{array}$ & 59.01390   & 53.876442  & $\begin{array}[t]{r}0.00404\\       -\end{array}$   & $\begin{array}[t]{r}303.7\\    -\end{array}$ & $\begin{array}[t]{r}  56\\    -\end{array}$ & S & N & GPS439230543517      & 50 & $\begin{array}[t]{r}13.75\\ 13.30\end{array}$ & $\begin{array}[t]{r}11.48\\11.01\end{array}$ & $\begin{array}[t]{r} 9.81\\  9.31\end{array}$ &  7.981 &  6.732 &  4.744 &  1.280 & Y & Y & Y & Y & N \\
$\begin{array}[t]{c}{\rm MHO}\\3722\end{array}$ & 59.14762   & 53.885530  & $\begin{array}[t]{r}0.00192\\       -\end{array}$   & $\begin{array}[t]{r}297.5\\    -\end{array}$ & $\begin{array}[t]{r}  22\\    -\end{array}$ & S & N & GPS439230543314      & 70 & $\begin{array}[t]{r}19.73\\     -\end{array}$ & $\begin{array}[t]{r}17.99\\    -\end{array}$ & $\begin{array}[t]{r}17.06\\     -\end{array}$ & 16.377 & 14.882 & 11.930 &  8.335 & Y & N & Y & N & N \\
$\begin{array}[t]{c}{\rm MHO}\\3723\end{array}$ & 59.10104   & 53.886691  & $\begin{array}[t]{r}0.00705\\ 0.00233\end{array}$   & $\begin{array}[t]{r} 74.5\\273.4\end{array}$ & $\begin{array}[t]{r}  37\\    7\end{array}$ & B & N & GPS439230543299      & 50 & $\begin{array}[t]{r}19.48\\     -\end{array}$ & $\begin{array}[t]{r}18.30\\    -\end{array}$ & $\begin{array}[t]{r}17.28\\     -\end{array}$ & 16.022 & 14.731 & 12.360 &  8.762 & Y & N & Y & N & N \\
$\begin{array}[t]{c}{\rm MHO}\\3724\end{array}$ & 59.06375   & 53.857752  & $\begin{array}[t]{r}0.00792\\ 0.00679\end{array}$   & $\begin{array}[t]{r} 45.3\\207.7\end{array}$ & $\begin{array}[t]{r} 171\\  106\end{array}$ & B & Y & GPS439230545538      & 50 & $\begin{array}[t]{r}    -\\     -\end{array}$ & $\begin{array}[t]{r}    -\\    -\end{array}$ & $\begin{array}[t]{r}16.84\\     -\end{array}$ & 13.337 & 11.395 &  7.320 &  3.217 & Y & N & Y & N & N \\
$\begin{array}[t]{c}{\rm MHO}\\3725\end{array}$ & 59.05799   & 53.856253  & $\begin{array}[t]{r}      -\\       -\end{array}$   & $\begin{array}[t]{r}    -\\    -\end{array}$ & $\begin{array}[t]{r}  47\\    -\end{array}$ & K & Y & unknown              &$-$ & $\begin{array}[t]{r}    -\\     -\end{array}$ & $\begin{array}[t]{r}    -\\    -\end{array}$ & $\begin{array}[t]{r}    -\\     -\end{array}$ &    $-$ &    $-$ &    $-$ &    $-$ & N & N & N & N & N \\
$\begin{array}[t]{c}{\rm MHO}\\3726\end{array}$ & 58.62680   & 53.381234  & $\begin{array}[t]{r}      -\\       -\end{array}$   & $\begin{array}[t]{r}    -\\    -\end{array}$ & $\begin{array}[t]{r}  71\\    -\end{array}$ & K & N & unknown              &$-$ & $\begin{array}[t]{r}    -\\     -\end{array}$ & $\begin{array}[t]{r}    -\\    -\end{array}$ & $\begin{array}[t]{r}    -\\     -\end{array}$ &    $-$ &    $-$ &    $-$ &    $-$ & N & N & N & N & N \\
$\begin{array}[t]{c}{\rm MHO}\\3727\end{array}$ & 58.23936   & 52.968608  & $\begin{array}[t]{r}0.00818\\       -\end{array}$   & $\begin{array}[t]{r}176.4\\    -\end{array}$ & $\begin{array}[t]{r}  25\\    -\end{array}$ & S & N & GPS439234689441      & 50 & $\begin{array}[t]{r}    -\\     -\end{array}$ & $\begin{array}[t]{r}18.52\\    -\end{array}$ & $\begin{array}[t]{r}16.69\\     -\end{array}$ & 14.396 & 11.795 &  9.003 &  5.446 & Y & N & Y & N & N \\
$\begin{array}[t]{c}{\rm MHO}\\3728\end{array}$ & 58.33467   & 52.930203  & $\begin{array}[t]{r}0.03869\\ 0.03221\end{array}$   & $\begin{array}[t]{r}  3.3\\184.3\end{array}$ & $\begin{array}[t]{r} 184\\  920\end{array}$ & B & N & noname               & 90 & $\begin{array}[t]{r}    -\\     -\end{array}$ & $\begin{array}[t]{r}    -\\    -\end{array}$ & $\begin{array}[t]{r}    -\\     -\end{array}$ &    $-$ &    $-$ &    $-$ &    $-$ & N & N & N & Y & N \\
$\begin{array}[t]{c}{\rm MHO}\\3729\end{array}$ & 58.80873   & 53.352021  & $\begin{array}[t]{r}0.00122\\ 0.00094\end{array}$   & $\begin{array}[t]{r} 67.9\\253.4\end{array}$ & $\begin{array}[t]{r}  26\\   42\end{array}$ & B & N & GPS439233238928      & 90 & $\begin{array}[t]{r}19.22\\ 17.06\end{array}$ & $\begin{array}[t]{r}17.52\\16.35\end{array}$ & $\begin{array}[t]{r}15.37\\ 15.00\end{array}$ & 12.991 & 13.321 &  6.096 &  2.686 & Y & Y & Y & Y & Y \\
$\begin{array}[t]{c}{\rm MHO}\\3730\end{array}$ & 58.71942   & 53.167113  & $\begin{array}[t]{r}0.00409\\ 0.00568\end{array}$   & $\begin{array}[t]{r}331.6\\141.0\end{array}$ & $\begin{array}[t]{r}  36\\   40\end{array}$ & B & N & GPS439233906954      & 50 & $\begin{array}[t]{r}    -\\     -\end{array}$ & $\begin{array}[t]{r}18.55\\    -\end{array}$ & $\begin{array}[t]{r}15.11\\     -\end{array}$ & 11.905 &  9.419 &  6.174 &  3.299 & Y & N & Y & Y & Y \\
$\begin{array}[t]{c}{\rm MHO}\\3731\end{array}$ & 58.71942   & 53.167113  & $\begin{array}[t]{r}0.05833\\ 0.07911\end{array}$   & $\begin{array}[t]{r} 48.2\\229.0\end{array}$ & $\begin{array}[t]{r} 281\\  557\end{array}$ & B & N & GPS439233906954      & 50 & $\begin{array}[t]{r}    -\\     -\end{array}$ & $\begin{array}[t]{r}18.55\\    -\end{array}$ & $\begin{array}[t]{r}15.11\\     -\end{array}$ & 11.905 &  9.419 &  6.174 &  3.299 & Y & N & Y & Y & Y \\
$\begin{array}[t]{c}{\rm MHO}\\3732\end{array}$ & 58.73051   & 53.170239  & $\begin{array}[t]{r}      -\\       -\end{array}$   & $\begin{array}[t]{r}    -\\    -\end{array}$ & $\begin{array}[t]{r}  31\\    -\end{array}$ & K & N & unknown              &$-$ & $\begin{array}[t]{r}    -\\     -\end{array}$ & $\begin{array}[t]{r}    -\\    -\end{array}$ & $\begin{array}[t]{r}    -\\     -\end{array}$ &    $-$ &    $-$ &    $-$ &    $-$ & N & N & N & N & N \\
$\begin{array}[t]{c}{\rm MHO}\\3733\end{array}$ & 58.71409   & 53.161405  & $\begin{array}[t]{r}      -\\       -\end{array}$   & $\begin{array}[t]{r}    -\\    -\end{array}$ & $\begin{array}[t]{r}  53\\    -\end{array}$ & K & N & unknown              &$-$ & $\begin{array}[t]{r}    -\\     -\end{array}$ & $\begin{array}[t]{r}    -\\    -\end{array}$ & $\begin{array}[t]{r}    -\\     -\end{array}$ &    $-$ &    $-$ &    $-$ &    $-$ & N & N & N & N & N \\
$\begin{array}[t]{c}{\rm MHO}\\3734\end{array}$ & 57.90289   & 51.484970  & $\begin{array}[t]{r}0.00472\\ 0.00758\end{array}$   & $\begin{array}[t]{r} 45.1\\220.1\end{array}$ & $\begin{array}[t]{r}  26\\   93\end{array}$ & B & Y & GPS439240289847      & 40 & $\begin{array}[t]{r}    -\\     -\end{array}$ & $\begin{array}[t]{r}    -\\    -\end{array}$ & $\begin{array}[t]{r}17.41\\     -\end{array}$ &    $-$ &    $-$ &    $-$ &    $-$ & Y & N & N & N & N \\
$\begin{array}[t]{c}{\rm MHO}\\3735\end{array}$ & 57.90289   & 51.484970  & $\begin{array}[t]{r}0.00121\\ 0.02414\end{array}$   & $\begin{array}[t]{r} 74.6\\267.1\end{array}$ & $\begin{array}[t]{r}  16\\   74\end{array}$ & B & Y & GPS439240289847      & 40 & $\begin{array}[t]{r}    -\\     -\end{array}$ & $\begin{array}[t]{r}    -\\    -\end{array}$ & $\begin{array}[t]{r}17.41\\     -\end{array}$ &    $-$ &    $-$ &    $-$ &    $-$ & Y & N & N & N & N \\
$\begin{array}[t]{c}{\rm MHO}\\3736\end{array}$ & 59.96436   & 52.554540  & $\begin{array}[t]{r}0.00480\\ 0.00340\end{array}$   & $\begin{array}[t]{r}156.4\\340.9\end{array}$ & $\begin{array}[t]{r} 141\\  124\end{array}$ & B & N & GPS439235701408      & 50 & $\begin{array}[t]{r}17.24\\ 16.59\end{array}$ & $\begin{array}[t]{r}15.84\\15.20\end{array}$ & $\begin{array}[t]{r}14.58\\ 13.36\end{array}$ & 12.980 & 10.904 &  7.912 &  4.006 & Y & Y & Y & Y & Y \\

\end{longtable}
\label{lastpage}\end{landscape}

\setlength\LTcapwidth{\textwidth}

\end{appendix}

\end{document}